\documentclass[useAMS,usenatbib]{mn2e}
\usepackage[usenames,dvipsnames]{color}

\newcommand{\Msun}{\ensuremath{\mathrm{M}_\odot}}

\newcommand{\red}[1]{{\color{red} #1} }

\newcommand{\Hp}{\ensuremath{H_{\mathrm{p}}}}
\newcommand{\fcbm}{\ensuremath{f_\mathrm{CBM}}}

\newcommand{\mesa}{\texttt{MESA}}
\newcommand{\template}{\textsc{\texttt{template}}}
\newcommand{\Ca}{\textsc{\texttt{C1}}}
\newcommand{\Cb}{\textsc{\texttt{C2}}}
\newcommand{\Cc}{\textsc{\texttt{C3}}}
\newcommand{\ONea}{\textsc{\texttt{ONe1}}}
\newcommand{\ONeb}{\textsc{\texttt{ONe2}}}
\newcommand{\ONec}{\textsc{\texttt{ONe3}}}
\newcommand{\Sia}{\textsc{\texttt{Si1}}}
\newcommand{\Sib}{\textsc{\texttt{Si2}}}
\newcommand{\Sic}{\textsc{\texttt{Si3}}}
\newcommand{\C}{\textsc{\texttt{C}}}
\newcommand{\ONe}{\textsc{\texttt{ONe}}}
\newcommand{\Si}{\textsc{\texttt{Si}}}

\newcommand{\nuclei}[2]{\ensuremath{\mathrm{^{#1}#2}}}

\newcommand{\helium}[1][4]{\nuclei{#1}{He}}
\newcommand{\carbon}[1][12]{\nuclei{#1}{C}}
\newcommand{\oxygen}[1][16]{\nuclei{#1}{O}}
\newcommand{\neon}[1][20]{\nuclei{#1}{Ne}}

\newcommand{\magnesium}[1][24]{\nuclei{#1}{Mg}}
\newcommand{\magnesiumB}[1][25]{\nuclei{#1}{Mg}}
\newcommand{\silicon}[1][28]{\nuclei{#1}{Si}}

\newcommand{\num}[1]{{#1}}

\usepackage{graphicx}
\graphicspath{{figures/}}
\usepackage{amssymb}
\usepackage{amsmath}
\usepackage{color}
\usepackage{soul}
\usepackage{booktabs}
\usepackage{afterpage}
\usepackage{url}
\usepackage{enumerate}

\title{Convective boundary mixing in a post-He core burning massive star model}

\author[A.~Davis et.~al]{
A.~Davis$^1$\thanks{E-mail: adavis@uvic.ca},
S.~Jones$^{2,3}$,
F.~Herwig$^{1,4}$,
\\
$^{1}$Department of Physics \& Astronomy, University of Victoria, P.O. Bos 3055
Victoria, B.C., V8W 3P6, Canada\\
$^{2}$Computational Physics and Methods (CCS-2), Los Alamos National
Laboratory, 87544 New Mexico, USA\\
$^{3}$Heidelberg Institute for Theoretical Studies, Schloss-Wolfsbrunnenweg 35,
D-69118 Heidelberg, Germany\\
$^{4}$Joint Institute for Nuclear Astrophysics, Center for the Evolution of the
Elements, Michigan State University, \\ 640 South Shaw Lane, East Lansing, MI
48824, USA}

\begin{document}



\maketitle

\label{firstpage}

\begin{abstract} Convective boundary mixing (CBM) in the advanced
	evolutionary stages of massive stars is not well understood. Structural
	changes caused by convection have an impact on the evolution as well as
	the subsequent supernova, or lack thereof. The effects of convectively 
	driven mixing across convective boundaries during the post He core burning 
	evolution of $25\Msun$, solar-metallicity, non-rotating stellar models is 
	studied using the \mesa~stellar evolution code. CBM is modelled using 
	the exponentially decaying diffusion coefficient 
	equation, the free parameter of which, \fcbm, is varied systematically throughout 
	the course of the stellar model's evolution with values of (0.002, 0.012, 
	0.022, 0.032). The effect of varying this 
	parameter produces mass ranges at collapse in the ONe, Si, Fe cores 
	of (1.82\Msun, 4.36\Msun), (1.67\Msun, 1.99\Msun) and (1.46\Msun, 
	1.70\Msun) respectively, with percent differences from the model with 
	minimal CBM as large as 86.3\%. At the presupernova stage, the 
	compactness of the stellar cores from \citet{OConnor2011}, $\xi_M$, exhibit a range 
	of (0.120, 0.354), suggesting that the extent 
	of CBM in the advanced burning stages of massive 
	stars is an important consideration for the explodability and type of 
	compact remnant. The nucleosynthetic yields from the models, most 
	notably C, O, Ne, Mg and Si are also significantly affected by the 
	CBM assumptions, showing non-linear trends 
	with increased mixing. The simulations show that interactions 
	between convective C, Ne and O shells produce significant non-linear 
	changes in the evolution, whereas from the end of Si burning, the 
	structural changes attributed to the CBM 
	are dominated by the growth of the convective C shell.
	Progenitor structures for all the models are available from \red{HERE
	(link and DOI to appear)}.

\end{abstract}


\section{Introduction} \label{sec:intro}

\begin{figure*}
	\includegraphics[width=\linewidth, clip=true, trim=0mm 2mm 0mm 0mm]{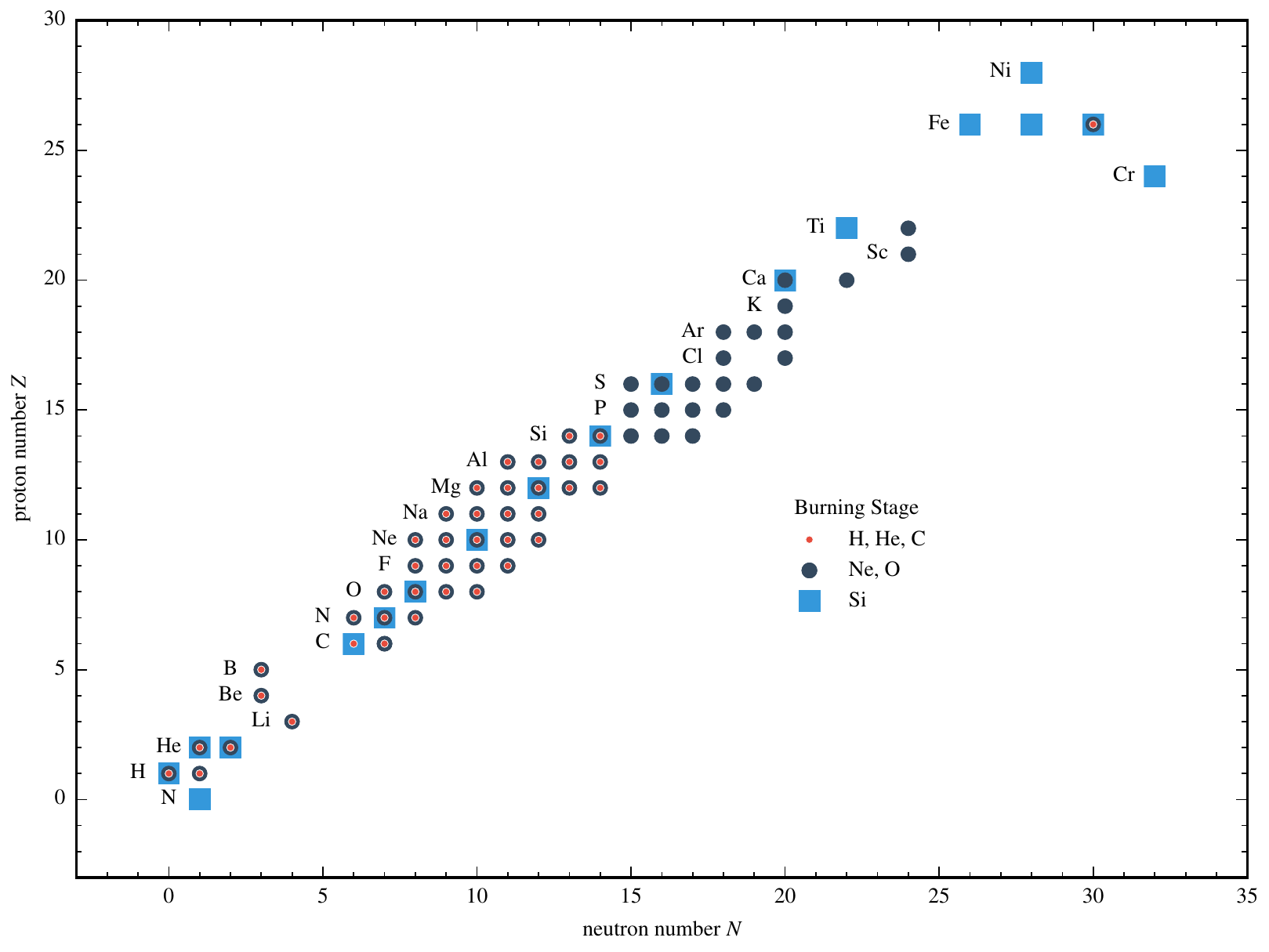}
	\caption{
		The nuclear reaction network used in the \mesa~
		simulations. The different marker styles represent the evolutionary 
		stage where the element was included in the reaction network.
	}
	\label{fig:network} 
\end{figure*}

\begin{figure*}
	\includegraphics[width=\linewidth, clip=true, trim=0mm 2mm 0mm 
	3mm]{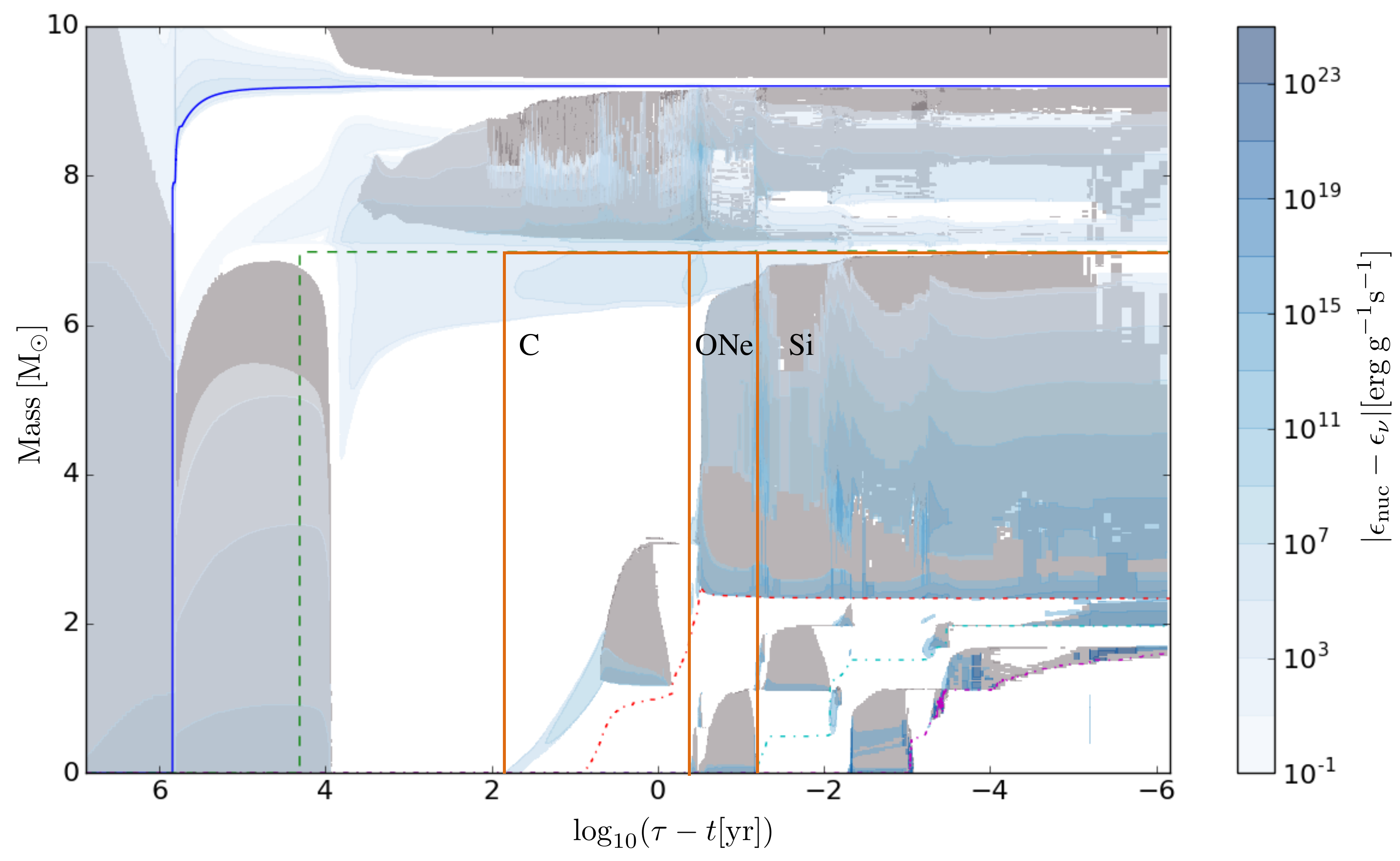}
	\caption{
		The regions of the \mesa~simulation under investigation. The
		Kippenhahn diagram is taken from the \template\ simulation.
		Grey areas represent regions that are convectively unstable and
		the blue contours are regions of positive energy generation
		where $\epsilon_\nu$ is the specific energy loss rate due to
		neutrino production. The x-axis is given in log of the time
		until the star collapses, where $\tau$ is the stars age at collapse. 
		The solid blue line marks the H free core ($X_{\mathrm{H}} < 10^{-2}$), 
		the dashed green line is the CO core boundary 
		($X_{\ensuremath{\mathrm{^{4}He}}} < 10^{-2}$),
		the dash-dotted red line is the ONe core
		($X_{\ensuremath{\mathrm{^{12}C}}} < 10^{-2}$), the dash-dotted
		light blue line is the Si core
		($X_{\ensuremath{\mathrm{^{16}O}}} < 10^{-2}$) and the
		dash-dotted magenta line is the Fe core
		($X_{\ensuremath{\mathrm{^{28}Si}}} < 10^{-2}$). The orange
		lines mark the region of interest for this study. The
		vertical orange lines mark the beginning of core burning
		stages which are the points where the
		CBM is increased for each of the respective run sets (\C, \ONe,
		\Si).
	}
	\label{fig:2_kip_runs} 
\end{figure*}

The predictions of stellar evolution are subject to a number of uncertainties, both from a modelling and physical perspective. The effect of rotation and magnetic fields on the evolution of massive stars have been studied extensively in the past \citep[e.g.][]{Heger:2005bi}. \cite{Farmer2016} analyzed the uncertainty introduced by different assumptions on the size of the nuclear network, which can have significant influence on the evolution leading up to the presupernova stage. More recently \citet{Sukhbold2017} investigated the effect of mass loss on the presupernova structure and explodability of a wide range of massive stars. Another significant source of uncertainty in stellar evolution is the treatment of convective boundaries. The effects that different convective boundary mixing (CBM) strengths have on the post-He core evolution of massive stars has not been investigated.
 
At the boundaries of convective regions in the star, fluid instabilities allow for the mixing of material between convectively stable and unstable regions. Simulations by 
\citet{Freytag1996,Meakin2007,Woodward:2013uf,Viallet2015,Jones2016} have shown that the mixing at convective boundaries is driven by the largest-scale turbulent convective flows. In low-mass stars, CBM has been shown to explain the observation of carbon
stars \citep{Herwig2000,Bertolli2013}. CBM may facilitate the interaction of convective shells in the
cores of massive stars \citep{Meakin2006,Jones2016}, and influence the light
curve and nucleosynthesis of novae \citep{Denissenkov2012}. \citet{Young2005}
showed that considering hydrodynamically-motivated mixing at convective
boundaries had a profound impact on the presupernova structure of massive
stars and, hence, their explosive nucleosynthesis. The influence of CBM on the
advanced burning stages of massive stars is the subject of this paper.

Massive stars are typically considered to be stars with a zero age main sequence
(ZAMS) mass of $M_\mathrm{ZAMS}\gtrsim8-9\Msun$ that end their lives as type-II or pair instability supernova. The lower limit is set by the
critical core mass for neon ignition
\citep{Nomoto1984,Jones2013,Jones2014,Doherty2015,Woosley2015}, which dependends on the initial metallicity of the star, extent of CBM of
the H and He burning cores \citep{Eldridge2004}, super-AGB star evolution \citep{Poelarends:2008fy} and rotation \citep{Farmer2015}. The internal structure of a massive star in its final evolutionary state consists of an inert core of Fe surrounded by shells burning Si, O, Ne, C, He and H 
\citep{Woosley2002}. These shells can be convectively unstable, and are then generally
separated by radiative regions composed of the ash from the shell
burning above.

Mixing mechanisms such as shearing, penetrative convection, gravity waves and
boundary layer separation, collectively referred to as CBM, act to mix material
from the stable layers into the convection zones and vice versa. Material mixed
from above into the convection zone may reach the deeper and hotter layers of the convection zone, producing regions of convective reactive nucleosynthesis \cite[e.g.][]{Herwig:2011dj,Ritter:2017vf}.
The core burning stages of a massive stars are sensitive to the CO core mass
left behind from He core burning, influencing the timing, extent
and luminosity, amongst other evolutionary characteristics 
\citep[e.g.][]{Sukhbold2014}. 

This paper presents the results of a numerical experiment examining the sensitivity of the
structure of one dimensional (1D) stellar evolution simulations of massive stars
with respect to CBM strength in the advanced burning stages (post-He core
burning evolution). Section~\ref{sec: methods} outlines the modelling
assumptions and Section~\ref{sec: results} describes the simulations and the key
findings, which are summarized and discussed in Section~\ref{sec: summary}.

\section{Methods} \label{sec: methods}

\begin{figure*}
	\includegraphics[width=\linewidth, clip=true, trim=0mm 2mm 0mm 3mm]{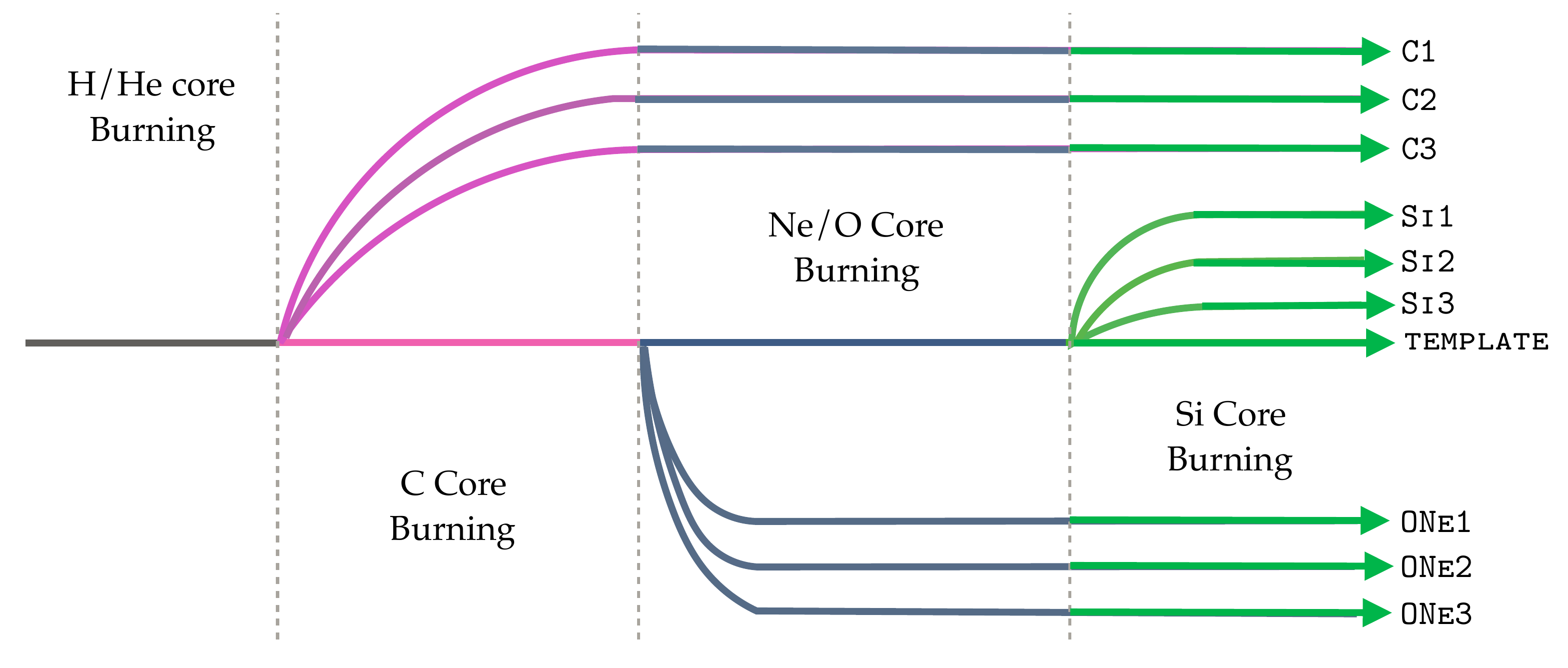}
	\caption{
		A diagram of the simulations computed for this study. Each solid line
		represents a simulation. The colours, separated by dashed lines, represent
		different core burning stages and the label at the end of each line is the run
		index (Table~\ref{tab:par}). The \template\ simulation was run to collapse
		with minimal \fcbm~(0.002) and all subsequent runs use a radial profile of the
		structure from the template as a staring position.
	}
	\label{fig:2_runs} 
\end{figure*}

Spherically symmetric (1D) stellar models with $M_\mathrm{ZAMS}=25\Msun$ and
initial metallicity $Z=0.02$ were computed using the \mesa~stellar
evolution code \citep{MESA2011,MESA2013,MESA2015}, revision 7184. The relative
abundances of the metals from \citet{Grevesse1993} was used.

\subsection{Input physics and modelling assumptions}

The \mesa~models were based on those from \citet{Jones2015}, to which a few
modifications were made.

\subsubsection{Mixing assumptions}

The Ledoux criterion \citep{Ledoux1947,Sakashita1959}
was used to determine convective stability instead of the Schwarzschild criterion\footnote{Comparing the \template\ simulation (Section~\ref{sec: methods-fcbm}) to the \mesa~results of \citet{Jones2015}, the CO core mass is larger with a percent difference of 1.61\% (Table~\ref{tab:par}), the \carbon/\oxygen\ at the end of He core burning is smaller at 0.334, with a percent difference of 7.22\%, and the main sequence and He core burning lifetimes were found to decrease to $6.911\mathrm{Myr}$ and $0.635\mathrm{Myr}$ with percent differences of 0.12\% and 2.5\% respectively.}
 \citep{Schwarzschild1958}, i.e. the influence of the mean molecular
weight $\mu$ on the stability was considered.  The effect of semi-convective
mixing (mixing on the thermal diffusion time scale in shells that have a
sub-adiabatic temperature gradient only owing to a stabilizing gradient in mean
molecular weight) is also considered, and uses the formulation of
\citet{Langer1983} who derived an effective diffusion coefficient from the
solution to Kato's equation for overstable, oscillatory convection in a
$\mu$-stratified medium \citep{Kato1966}. The efficiency parameter of
semiconvection, with which the corresponding effective diffusion coefficient is
multiplied, is taken to be $\alpha_{semi}=0.1$ \citep{Langer1983}, and is
considered to be fast semiconvection \citep[][see also Section 6.2 of \citealp{Maeder2009} for a discussion about the fidelity of this
approximation]{Woosley2007}. 

CBM is taken into account by the exponentially decaying diffusion model 
\citep{Freytag1996,Herwig1997}. In this model, the diffusion coefficient across
the convective boundary and into the formally stable layer is given by
\begin{equation}
\label{eq: DCBM}
D_{\mathrm{CBM}}(r)=D(r_0)
\exp\left\{{\dfrac{-2|r-r_0|}{f_{\mathrm{CBM}}\Hp(r_0)}}\right\}
\end{equation}
where $r_0$ is the radius of the convective boundary, $\Hp = dr/d\ln P$ is the
pressure scale height, and $D(r_0)$ is the diffusion
coefficient at the convective boundary given by $D(r_0)=\frac{1}{3}v_\mathrm{MLT}(r_0) \alpha_{\mathrm{MLT}} \Hp(r_0)$.
In this expresion $v_\mathrm{MLT}$ is the convective velocity and $\alpha_{\mathrm{MLT}}$ is the mixing length parameter. \fcbm~is the free parameter of
the model and the parameter of interest for this study. In \mesa~revision 7184, \fcbm\ can be specified for H, He and metal burning convection zones. A metal burning convection zone is
considered to be a convection zone whose peak nuclear energy generation
does not come from H or He burning. The consequences of varying the \fcbm\ parameter for the metal burning convection zones is the focus of this study (see Section~\ref{sec: methods-fcbm}).

\subsubsection{Nuclear reaction network}

A smaller nuclear reaction network is used in this work compared to
\citet{Jones2015}, who used a network consisting of 171 isotopes. This work
uses a moderate-sized reaction network from H to Si up until the start of Ne 
and O burning, where the network is extended through the alpha-elements up to
Fe (see Figure~\ref{fig:network}). A larger network is used for core Ne and O
burning than main sequence and He core burning to more accurately capture the
nuclear reactions involving heavier species found there.  For core Si burning,
the network is reduced to follow just 21 species to ease the computational
burden. While the 21-species network \texttt{approx21} explicitly tracks the
abundances of 21 species, $(\alpha,p)$ and $(p,\gamma)$ reactions (and their
inverses) are also included by assuming a steady-state of the intermediate
isotopes \citep{weaver1978}. One of the down-sides of switching networks is
that outside of the Si burning shells the electron fraction $Y_\mathrm{e}$ will
be reset to 0.5 because neutron excesses are only introduced via burning of Si
into Fe-group elements and electron captures by $^{56}$Ni and $^{56}$Fe. The
(de-)leptonization rate in the Fe core is set by the electron capture rates by
$^{56}$Ni and free protons and beta-decay and positron-capture by free neutrons
\citep[using reaction rates from][]{Langanke2000}. \citet{Farmer2016} have
investigated the effects of network size on presupernova structure which gives 
an idea of the uncertainty in the present work with respect to the reaction network. 

\subsection{Outline of the \fcbm~parameter study}\label{sec: methods-fcbm}

Following the extinction of convective core He burning\footnote{The end of the
core He burning phase is taken to be when the central He mass fraction falls
below $10^{-5}$.}, the model was evolved to the onset of iron core
collapse\footnote{The onset of core collapse was defined to be when the infall
velocity of the core reaches 1000~km~s$^{-1}$} using $\fcbm=0.002$ for the metal burning convection zones. This simulation is referred to throughout this manuscript as the \template\ simulation.

The \template\ simulation was then branched at three evolutionary stages
(as indicated in Figure~\ref{fig:2_kip_runs}):
\begin{enumerate}[(i)]
	\item the ignition of central C burning (\C\ series), when
		the central temperature exceeds $T_\mathrm{C}\simeq\num{7.59\times10^{8}}$K
	\item the ignition of central Ne burning (\ONe\ series), when
		the central temperature exceeds $T_\mathrm{C}\simeq\num{1.41\times10^{9}}$K
	\item the extinction of core O burning (\Si\ series), when the
		central temperature exceeds $T_\mathrm{C}\simeq\num{2.45\times10^{9}}$K.
\end{enumerate}
At each branching point, three models were generated, each assuming a different
value for \fcbm~(0.012, 0.022, 0.032), which was held constant across all
of the convection zones inside the CO core. Figure~\ref{fig:2_runs} gives a schematic diagram of the branching points for each simulation and Table~\ref{tab:par} gives the \fcbm~values for each burning stage of each simulation. The resolution of the resulting simulations consisted of $\approx 3000$ spatial points during the post He burning core evolution and $\approx 300,000$ temporal point over the lifetime.

\begin{table*}
	\centering
	\caption{Selected values for each simulation. The columns labeled \fcbm~
		followed by a core burning stage, are the CBM parameters implemented during
		that burning stage. The column labeled $M_{\mathrm{CO}}$ is the CO core mass at
		the end of He core burning and is the same for each simulation. $M_{\mathrm{ONe}}$ 
		is the mass of the ONe core when the simulations begin to burn Ne in the core 
		(Section \ref{sec: methods-fcbm}). $M_{\mathrm{Si}}$ is the mass of the Si core when convection 
		stops in the convective Si core. $M_{{\mathrm{Fe}}}$ is
		the mass of the Fe core at $\mathrm{log}_{10}(\tau - t) = -6$. The dashes in the core mass
		columns represent values that are the same as the \template~simulation. The bottom row 
		contains the range which is defined to be the absolute difference between the 
		highest and lowest values. Also included is the 
		percent differences (\% diff) for the smallest and largest core mass values compared to the 
		\template~simulation. Calculating a percent difference with the \template~simulation 
		is not meant to imply that the \template~value is the accepted value.
	}
	\begin{tabular}{lcccccccc}
		\toprule
		Name & 
		\fcbm (H, He) &
		\fcbm (C)&
		\fcbm (Ne, O)&
		\fcbm (Si)&
		$M_{\mathrm{CO}} [\Msun]$ &
		$M_{\mathrm{ONe}} [\Msun]$ &
		$M_{\mathrm{Si}} [\Msun]$ &
		$M_{\mathrm{Fe}} [\Msun]$

		\tabularnewline
		\midrule
		\template  &  0.002  &  0.002  &  0.002  &  0.002  &  6.93  &  1.77  &  1.51  &  1.59 \\
		\Ca        &  0.002  &  0.012  &  0.012  &  0.012  &  -     &  1.74  &  1.41  &  1.47 \\
		\Cb        &  0.002  &  0.022  &  0.022  &  0.022  &  -     &  1.77  &  1.60  &  1.56 \\
		\Cc        &  0.002  &  0.032  &  0.032  &  0.032  &  -     &  1.86  &  1.82  &  1.60 \\
		\ONea      &  0.002  &  0.002  &  0.012  &  0.012  &  -     &  -     &  1.46  &  1.46 \\
		\ONeb      &  0.002  &  0.002  &  0.022  &  0.022  &  -     &  -     &  1.61  &  1.54 \\
		\ONec      &  0.002  &  0.002  &  0.032  &  0.032  &  -     &  -     &  1.68  &  1.52 \\
		\Sia       &  0.002  &  0.002  &  0.002  &  0.012  &  -     &  -     &  -     &  1.62 \\
		\Sib       &  0.002  &  0.002  &  0.002  &  0.022  &  -     &  -     &  -     &  1.54 \\
		\Sic       &  0.002  &  0.002  &  0.002  &  0.032  &  -     &  -     &  -     &  1.43 \\
		\midrule
		           &         &         &         &         & range [\Msun]: &  0.12  &  0.41  &  0.19 \\
                           &         &         &         &         &  \% diff   &  (1.96, 5.08)  &  (6.62, 20.5)  &  (10.1, 1.89)  \\
		\bottomrule
	\end{tabular}
	\label{tab:par}
\end{table*}

The values of \fcbm~were chosen to span CBM strengths ranging from the lowest
value such that the simulation would converge without numerical smoothing
($\fcbm=0.002$), to a large value of $\fcbm=0.032$. The value of $\fcbm =
0.032$ has been deduced from idealized high-resolution 3D hydrodynamic
simulations of an O burning shell in a 25\Msun~star at the upper convective
boundary \citep{Jones2016}. Note, however, that the implementation presented
here is for the upper \emph{and} lower convective boundaries. Additionally, in
the analysis of \citet{Jones2016}, the effective diffusion coefficient was
decayed from a distance $\fcbm H_{\mathrm{P}}$ inside of the Schwarzschild
convective boundary. This parameter is generally denoted by $f_0$ and
represents a linear shift of the CBM model, from the convective boundary, into
the convection zone. In this study the value of $f_{0} = 0.002$ for all
simulations.  The value of $f_0$ was fixed in order to only test the effect of
\fcbm\ on the simulations.

Currently there is no convincing model for how CBM should depend on the
physical properties of the plasma (e.g. its thermodynamic state) and the flow
characteristics \citep[see][for the current status in CBM
modelling for stellar evolution]{Arnett2015,Viallet2015}.

Additionally, applying different values of \fcbm~to each type of convection 
zone (C burning, Ne burning, etc.), would significantly increase the 
parameter space of the study (number of \fcbm~values to the fourth 
power for C, Ne, O and Si), which is not the intention of this work
\footnote{However,
	let it be noted that a number of similar uncertainty studies of a more
	statistical nature have recently been published
	\citep{Farmer2015,Fields2016,Farmer2016}, the latter of which is
	concerned with the presupernova structure of massive stars and its
	sensitivity to numerical resolution and nuclear reaction network
size.}.  In view of the considerable uncertainties when adopting the
\fcbm~parameter, this numerical experiment is primarily designed to study the
sensitivity of the \mesa~simulations' stellar structure with respect to CBM
strength, at different times in the later stages of evolution.


\section{Results} \label{sec: results}

In this section, the effect of varying the \fcbm~parameter in the post He core
burning phases of a 25\Msun~stellar model is examined. The implications for
core structure, presupernova compactness and nucleosynthesis are presented. 
More detailed information on the structure of each simulation can be found in 
\citet{Davisthesis}\footnote{\url{http://hdl.handle.net/1828/8054}}.

\subsection{Convective structure} \label{sec: Convective structure}

\begin{figure}
	\includegraphics[width=\linewidth, clip=true, trim=0mm 2mm 0mm 3mm]{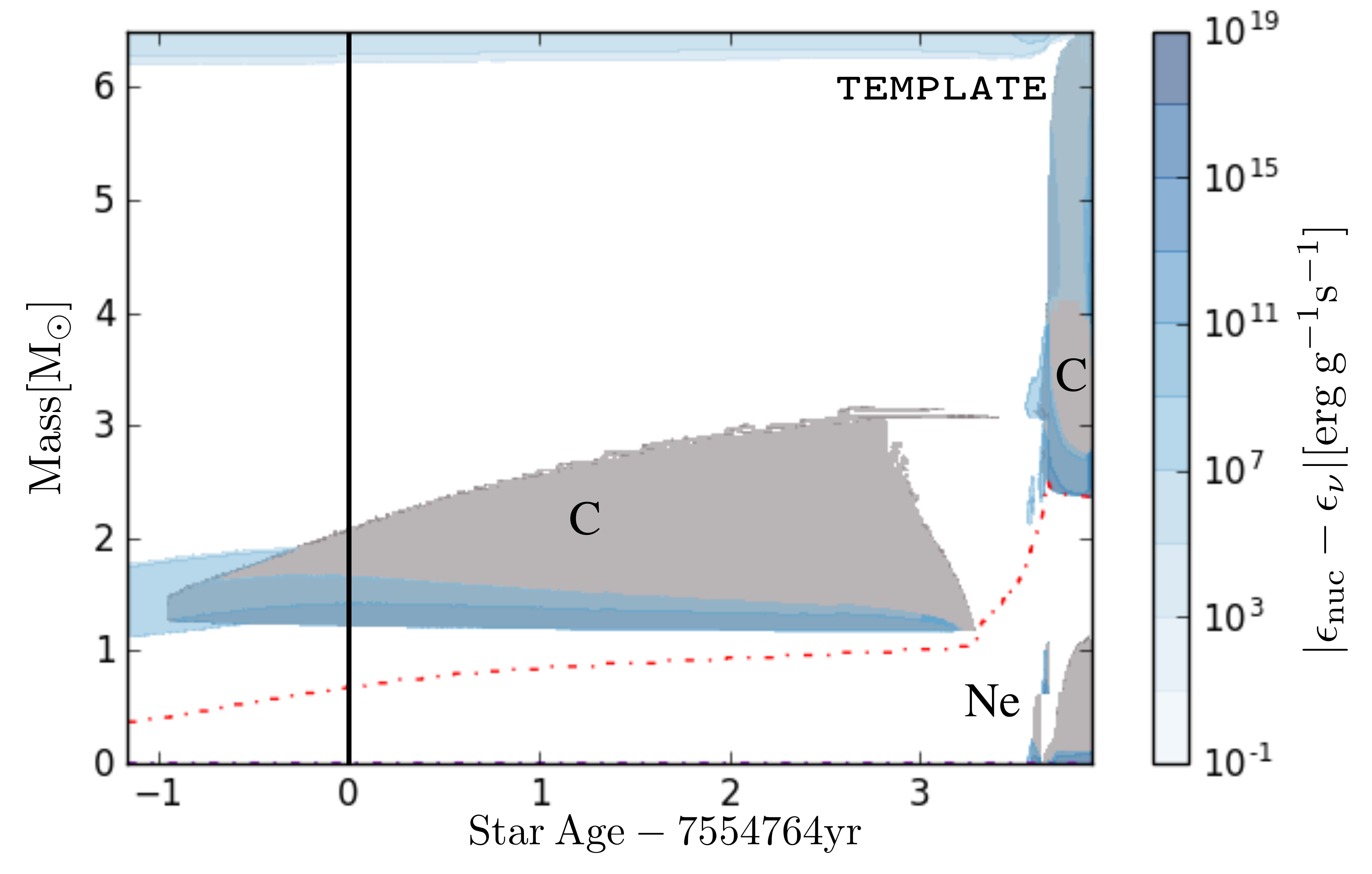}
	\includegraphics[width=\linewidth, clip=true, trim=0mm 2mm 0mm 3mm]{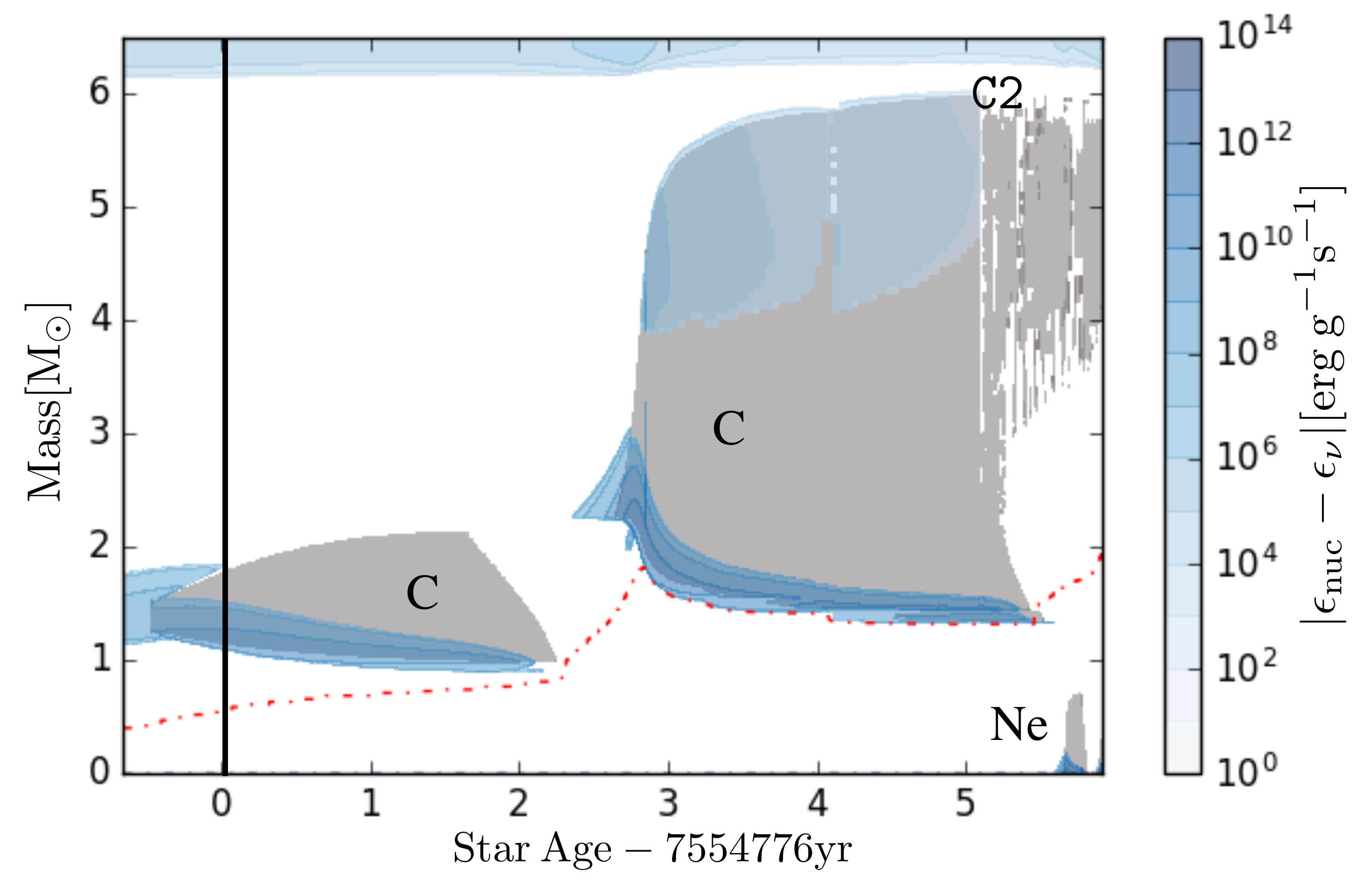}
	\includegraphics[width=\linewidth, clip=true, trim=0mm 2mm 0mm 3mm]{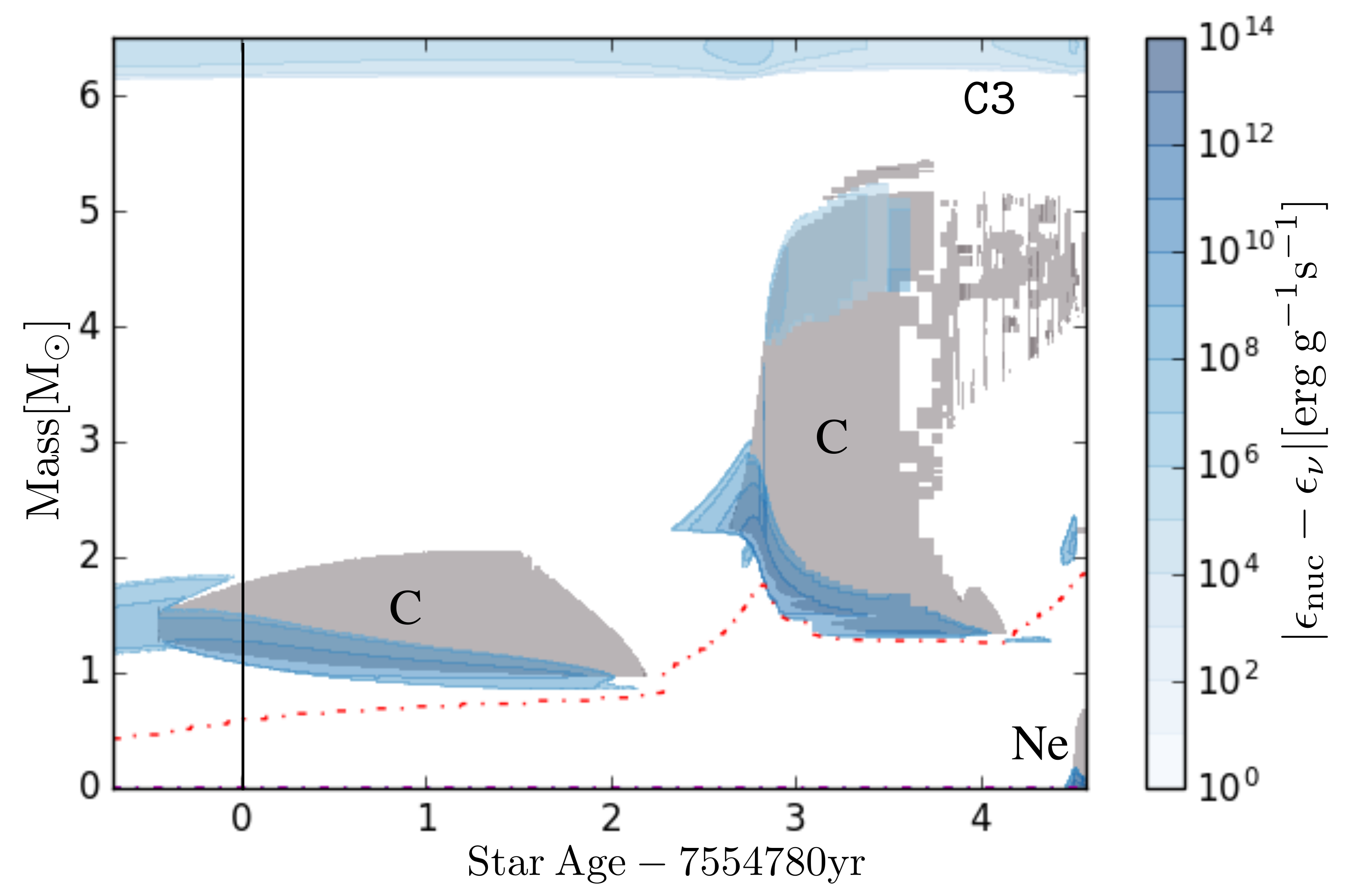}
	\caption{
	Kippenhahn diagrams for the \C\ simulations with increasing values 
	of \fcbm. The diagrams show the evolution of the first C shell up to 
	the beginning of the convective Ne core. The x-axis is the time from 
	the point of maximum luminosity due to C burning, $L_C$, in the 
	convective C shell.
	}
	\label{fig:3.1.1_kip} 
\end{figure}

\begin{figure}
	\includegraphics[width=\linewidth, clip=true, trim=0mm 5mm 0mm 0mm]{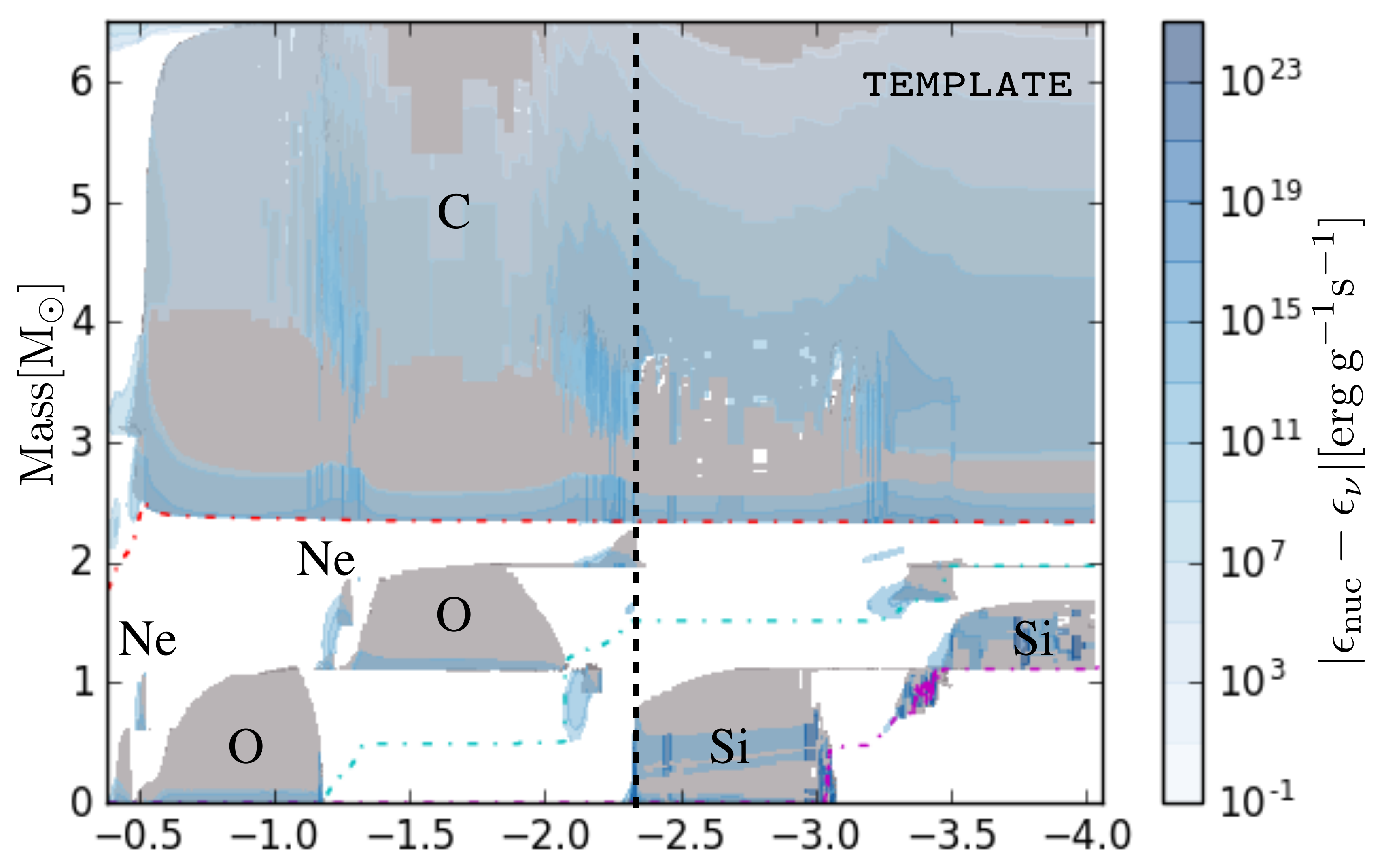}
	\includegraphics[width=\linewidth, clip=true, trim=0mm 5mm 0mm 0mm]{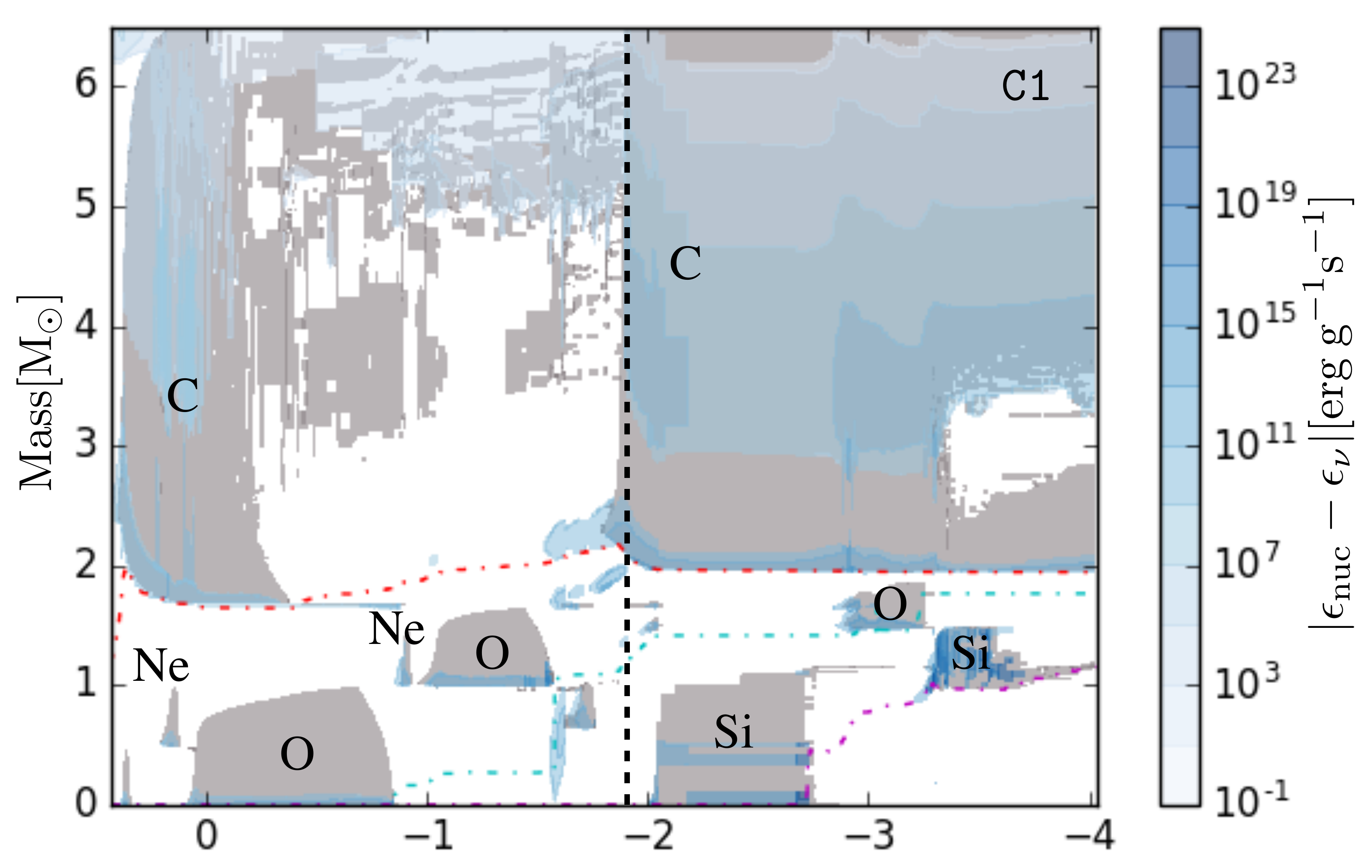}
	\includegraphics[width=\linewidth, clip=true, trim=0mm 5mm 0mm 0mm]{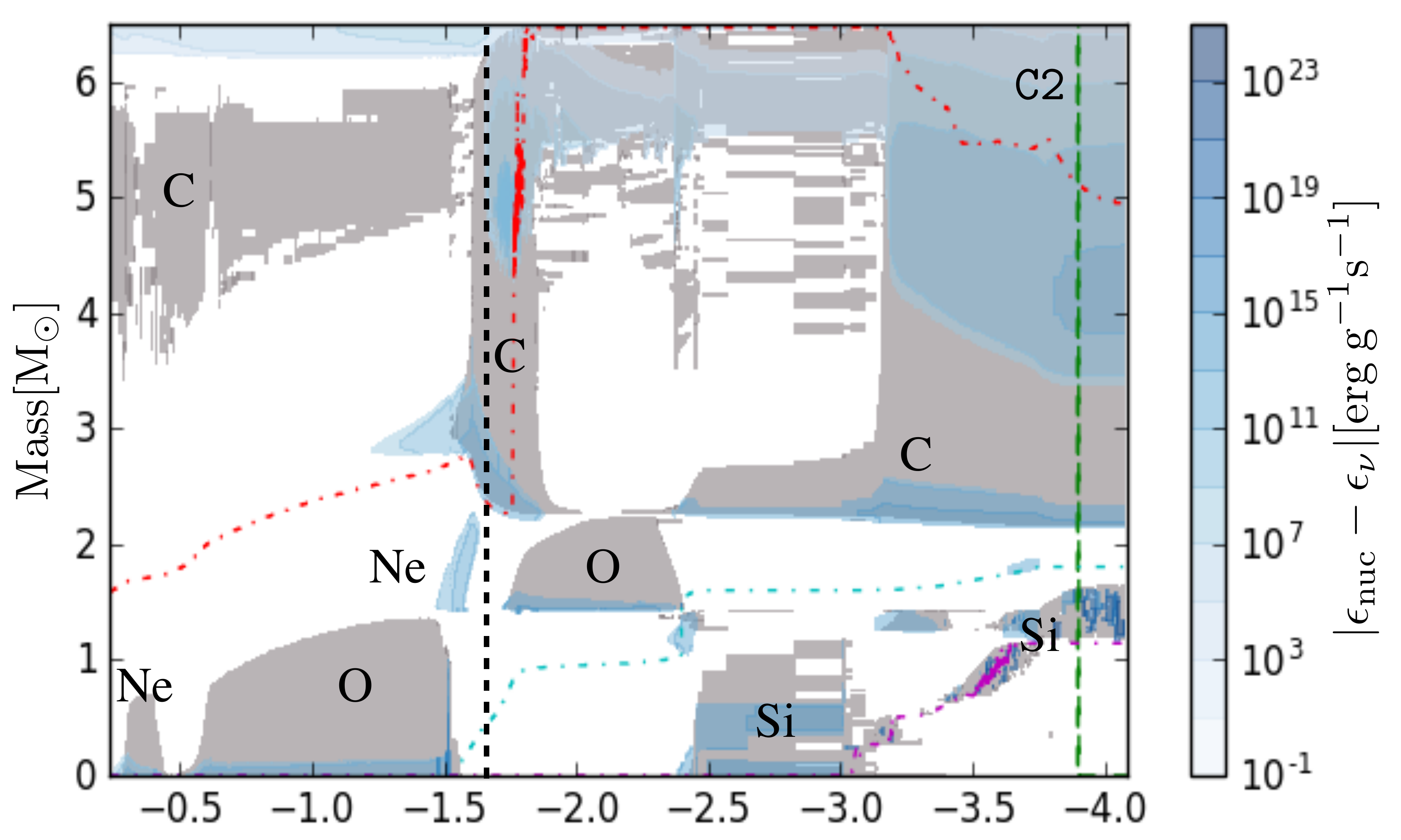}
	\includegraphics[width=\linewidth, clip=true, trim=0mm 0mm 5mm 5mm]{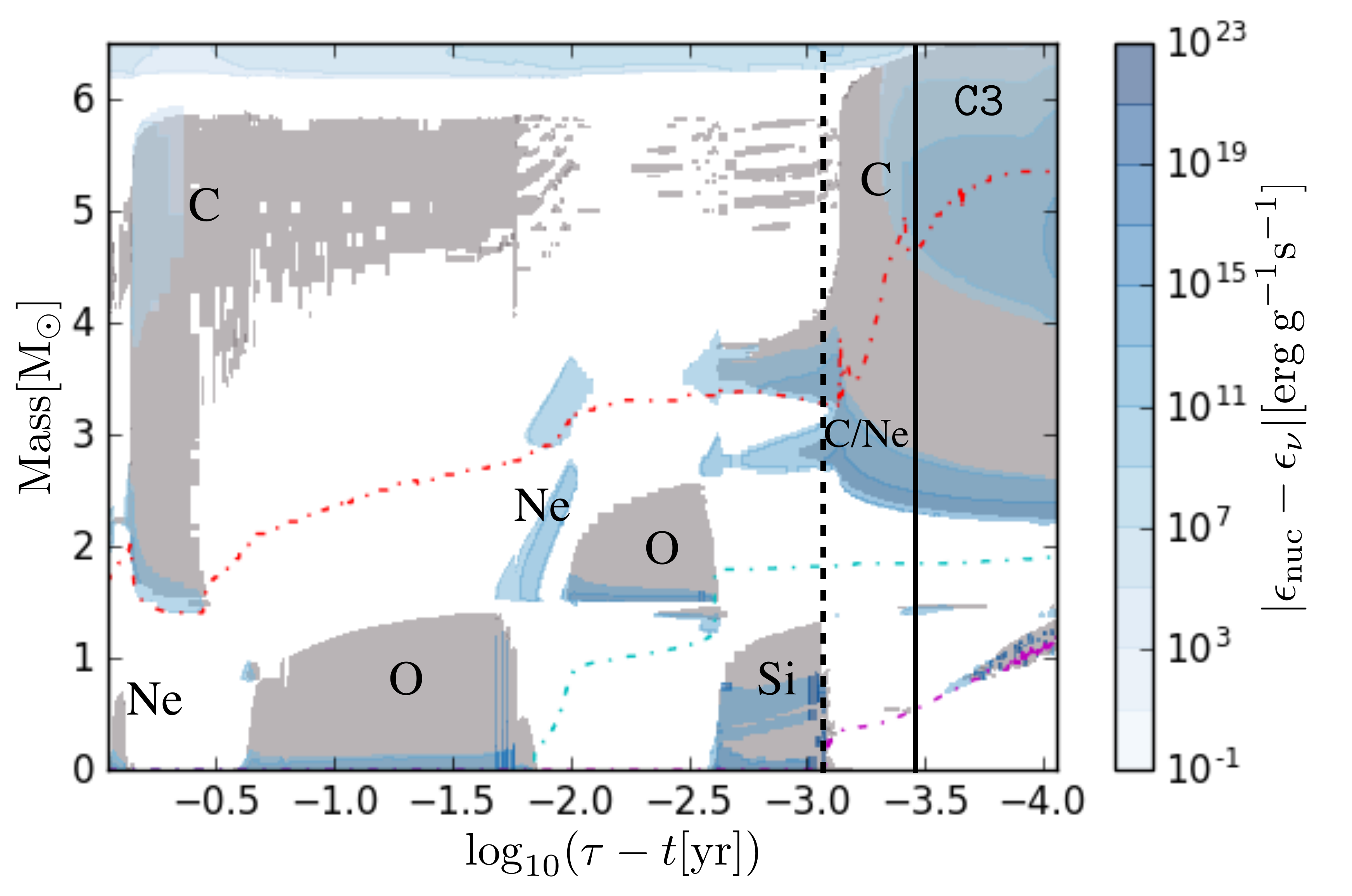}
	\caption{
		Kippenhahn diagrams for the \C\ simulations with increasing values 
		of \fcbm. The diagrams show the evolution from Ne core burning to
		$\approx 1\mathrm{hr}$ before collapse. The evolution here is a 
		continuation from Figure~\ref{fig:3.1.1_kip}. The dashed black lines 
		refer to Figure~\ref{fig:3.1.2_kip_abu}, \ref{fig:3.1.2_kip_abu_C32}, 
		\ref{fig:3.1.2_kip_s} and \ref{fig:3.1.2_kip_abu_C12}, and the solid 
		black lines refers to Figure~\ref{fig:3.1.2_L_abu}. Definitions for the 
		core boundaries can be found in Figure~\ref{fig:2_kip_runs}.
	}
	\label{fig:3.1.2_kip} 
\end{figure}

\begin{figure}
	\includegraphics[width=\linewidth, clip=true, trim=0mm 0mm 0mm 0mm]{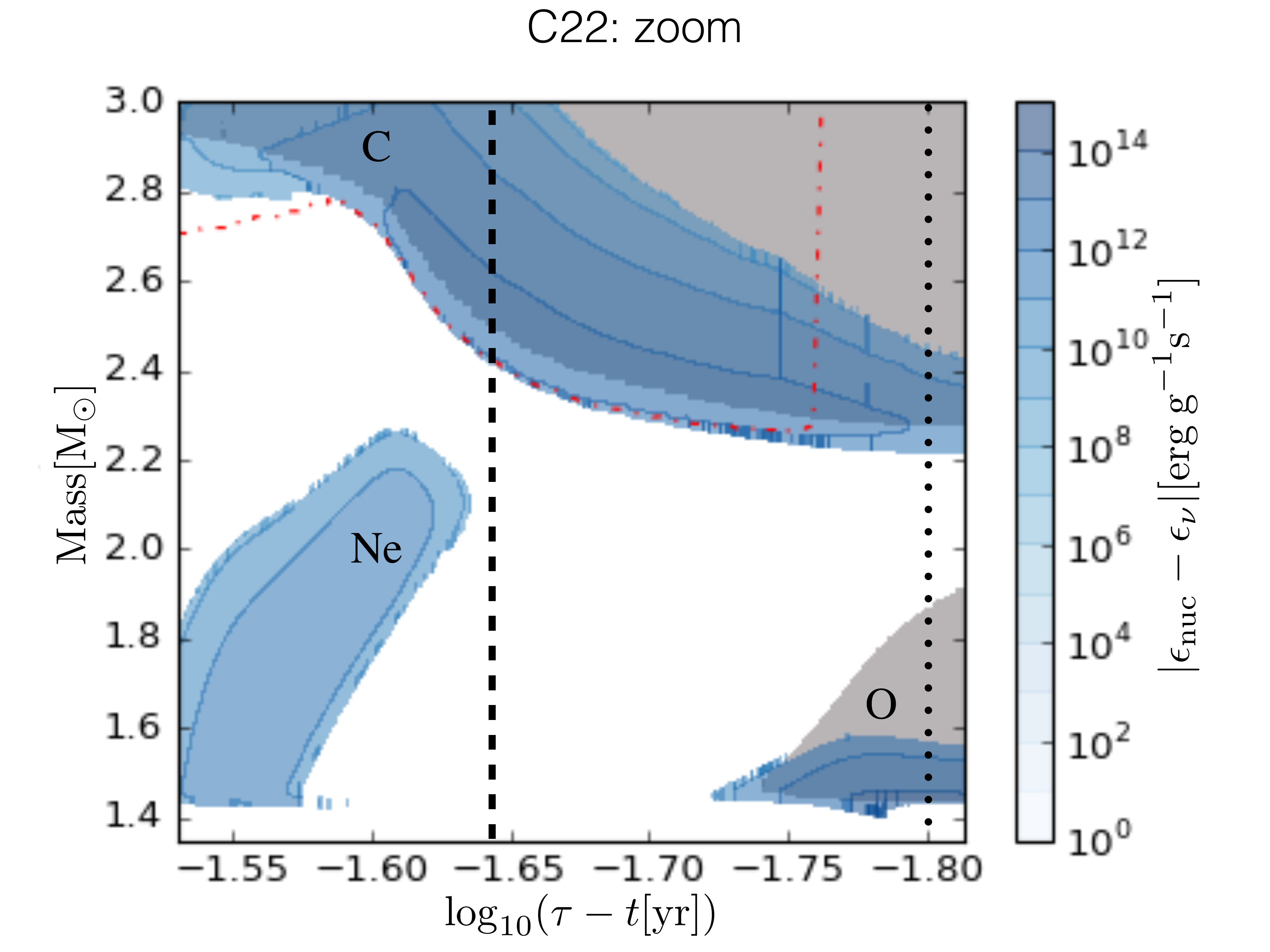}
	\includegraphics[width=\linewidth, clip=true, trim=0mm 0mm 0mm 0mm]{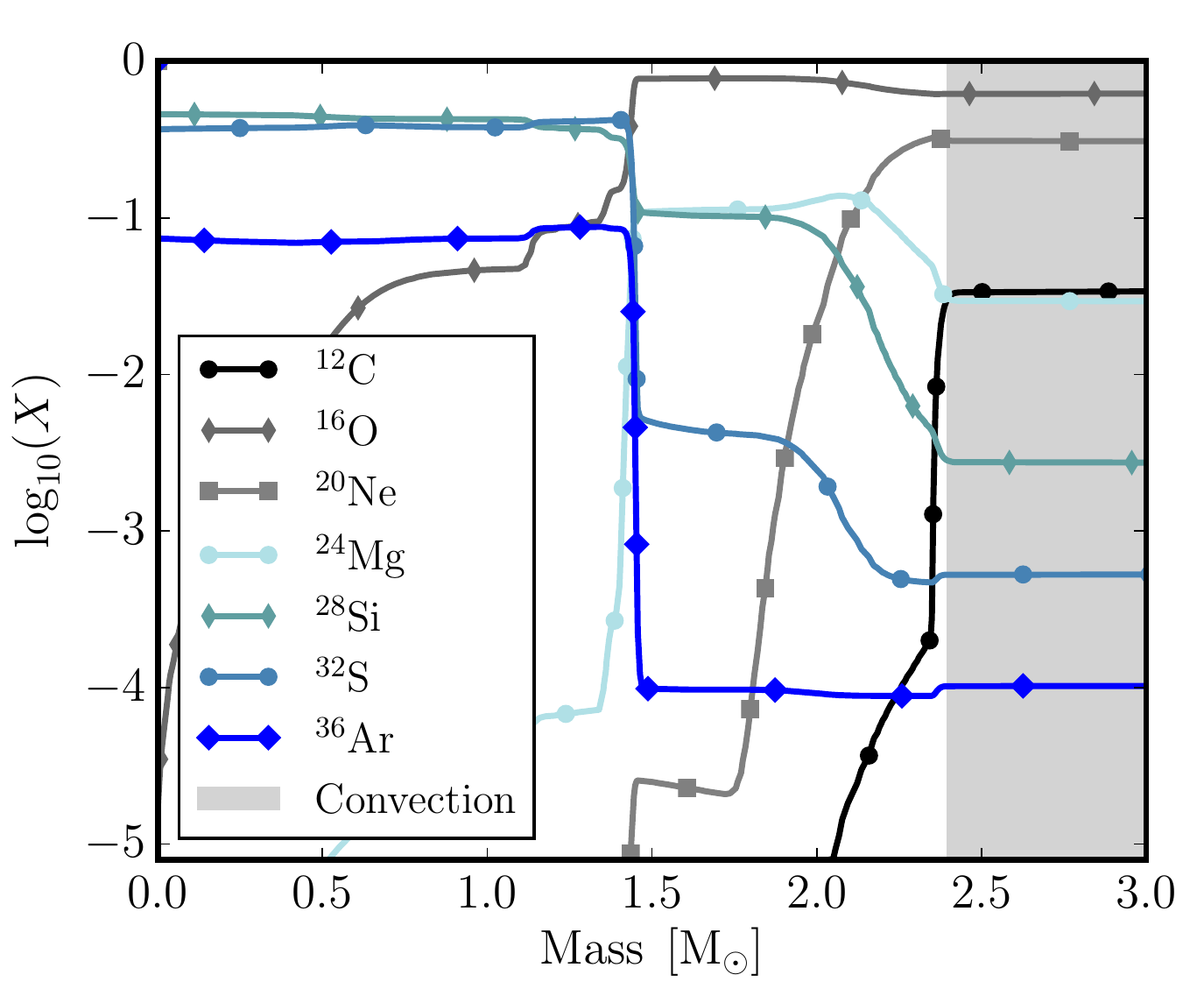}
	\caption{
		Kippenhahn diagram and abundance profile from the \Cb\ 
		simulation as the third C shell bottom boundary approaches 
		the Ne burning ash below. The upper convection 
		zone in the figure is the bottom of the C shell, a radiative Ne 
		burning region is on the left and the convection zone on the bottom 
		right is an O shell. The thin dash-dotted red line is the ONe 
		core boundary. The dotted black line is the point where the 
		convective boundary reaches it deepest in mass. The dashed 
		black line is the point where the abundance profile is taken. 
		This is the same dashed black line as in 
		Figure~\ref{fig:3.1.2_kip}. The abundance profile shows 
		the bottom boundary of the convective C shell mixing in Ne 
		ash from below (grey shaded area).
	}
	\label{fig:3.1.2_kip_abu} 
\end{figure}

\begin{figure}
	\includegraphics[width=\linewidth, clip=true, trim=0mm 0mm 0mm 0mm]{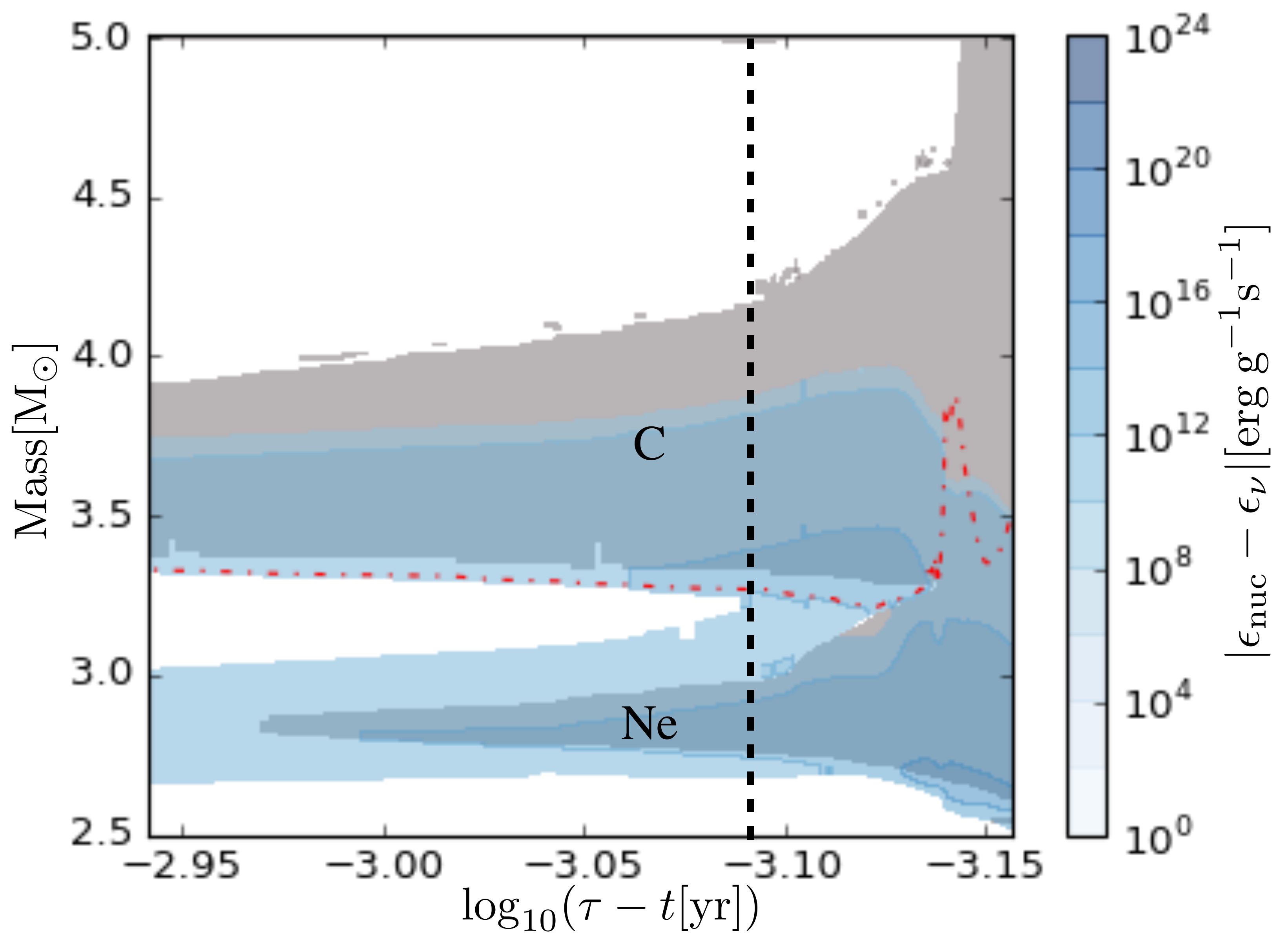}
	\includegraphics[width=\linewidth, clip=true, trim=0mm 0mm 0mm 0mm]{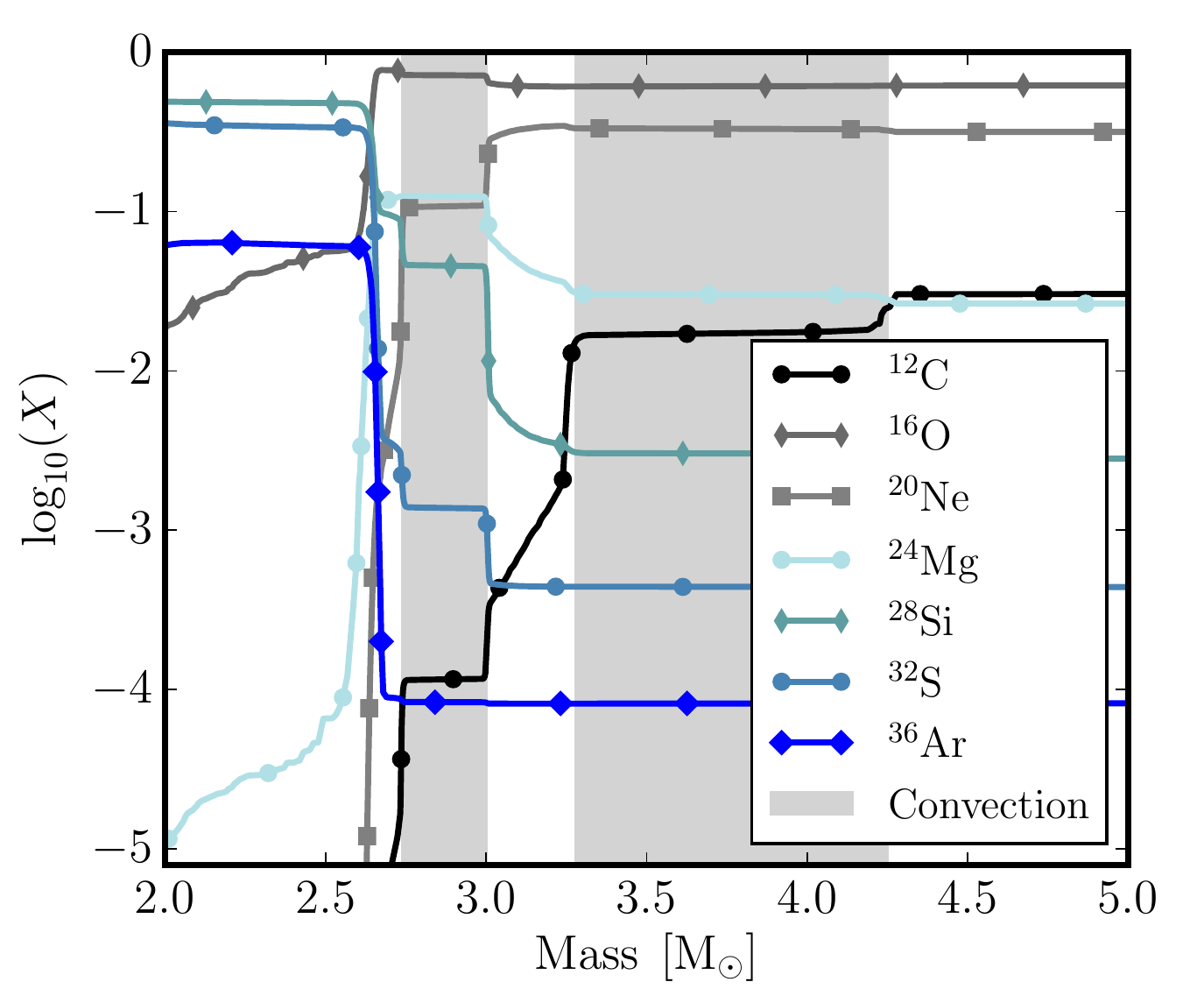}
	\caption{
		Kippenhahn diagram and abundance profile from the \Cc\ simulation. 
		The Kippenhahn diagram shows the merger of the Ne (bottom) and 
		C (top) shells. The red dash-dotted line is the ONe core boundary, 
		and the dashed black line is the point where the abundance profile is 
		taken and is the same dashed line as in Figure~\ref{fig:3.1.2_kip}. The 
		grey shaded regions on the abundance profile are the convection 
		zones in the Kippenhahn diagram.
	}
	\label{fig:3.1.2_kip_abu_C32} 
\end{figure}

\begin{figure}
	\includegraphics[width=\linewidth, clip=true, trim=0mm 0mm 0mm 0mm]{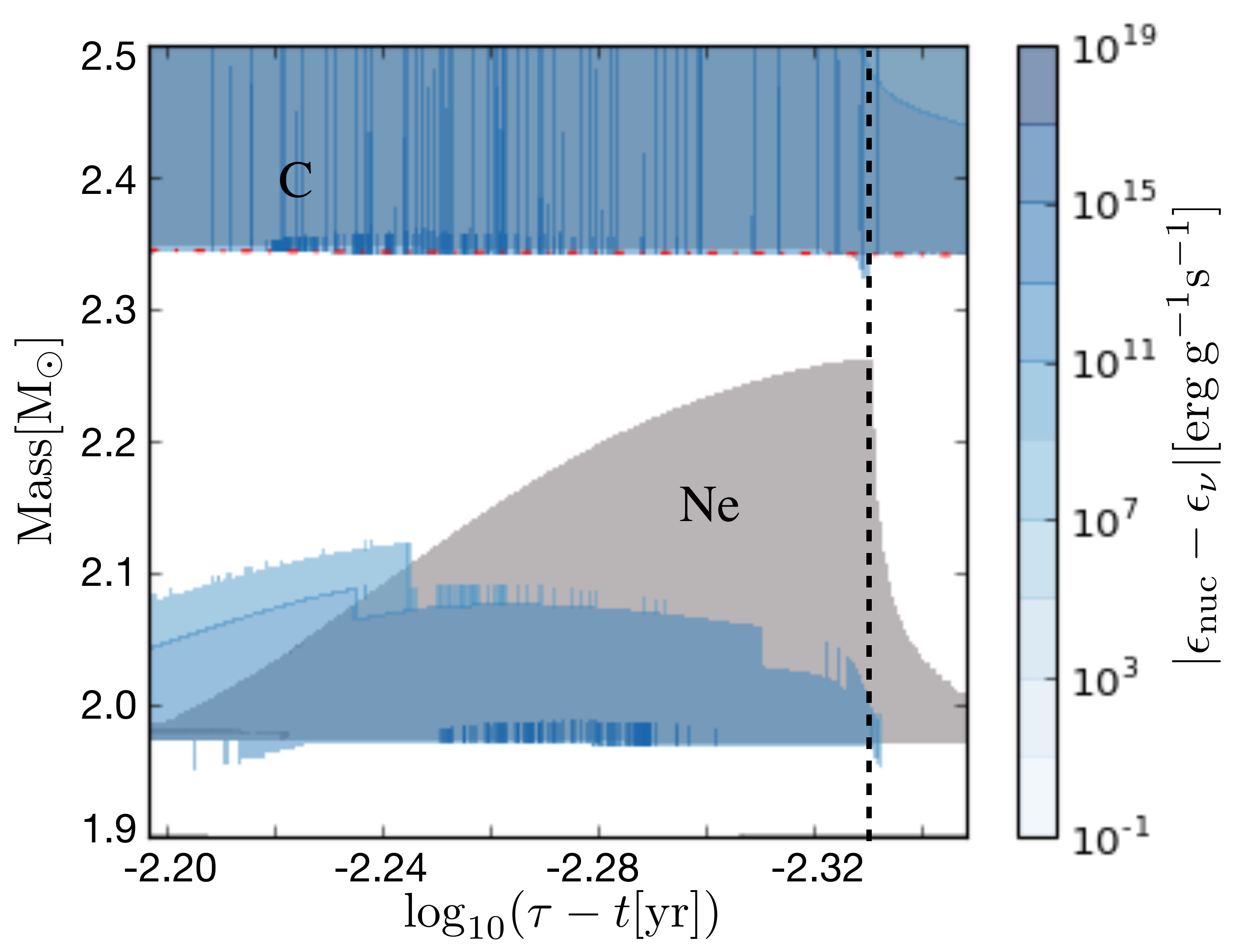}
	\includegraphics[width=\linewidth, clip=true, trim=0mm 0mm 0mm 0mm]{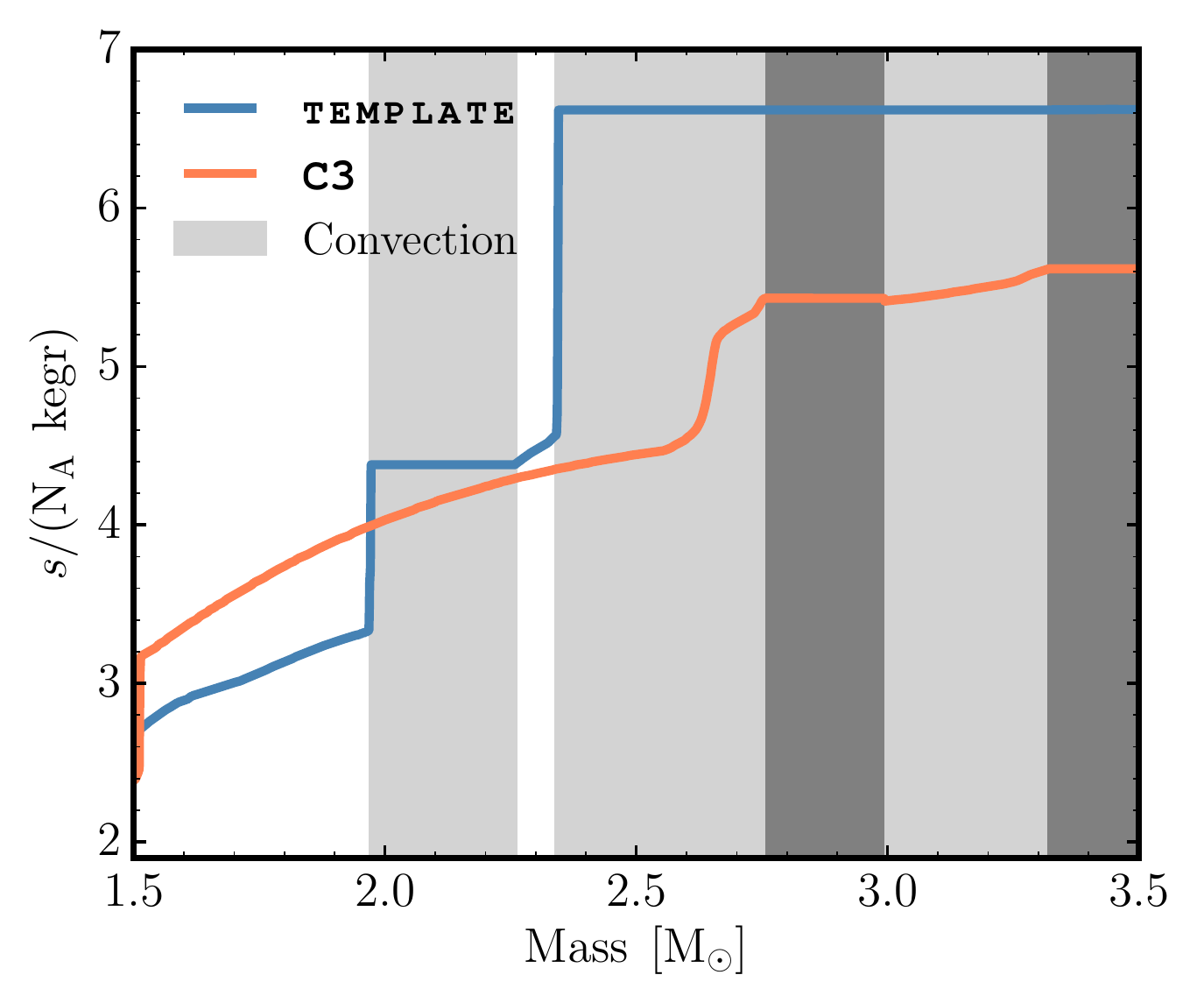}
	\caption{
		Kippenhahn diagram from the \template~simulation and entropy 
		profiles from the \template~and \Cc~simulations. The Kippenhahn 
		diagram shows a Ne shell underneath a convective C shell. The 
		entropy profile shows the large entropy gradient between the two
		convection zones in the \template~simulation and is taken at the 
		dashed black line in the Kippenhahn diagram (same dashed black 
		line as in Figure~\ref{fig:3.1.2_kip}). In the \Cc\ simulation, 
		the entropy gradient is much lower before the shell merger. This 
		profile is taken at the dashed black line in Figure~\ref{fig:3.1.2_kip} 
		for the \Cc\ simulation. Convective regions in the entropy plot are 
		illustrated by light grey shading for the \template\ and dark grey 
		for the \Cc\ simulation.
	}
	\label{fig:3.1.2_kip_s} 
\end{figure}

\begin{figure}
	\includegraphics[width=\linewidth, clip=true, trim=0mm 0mm 0mm 0mm]{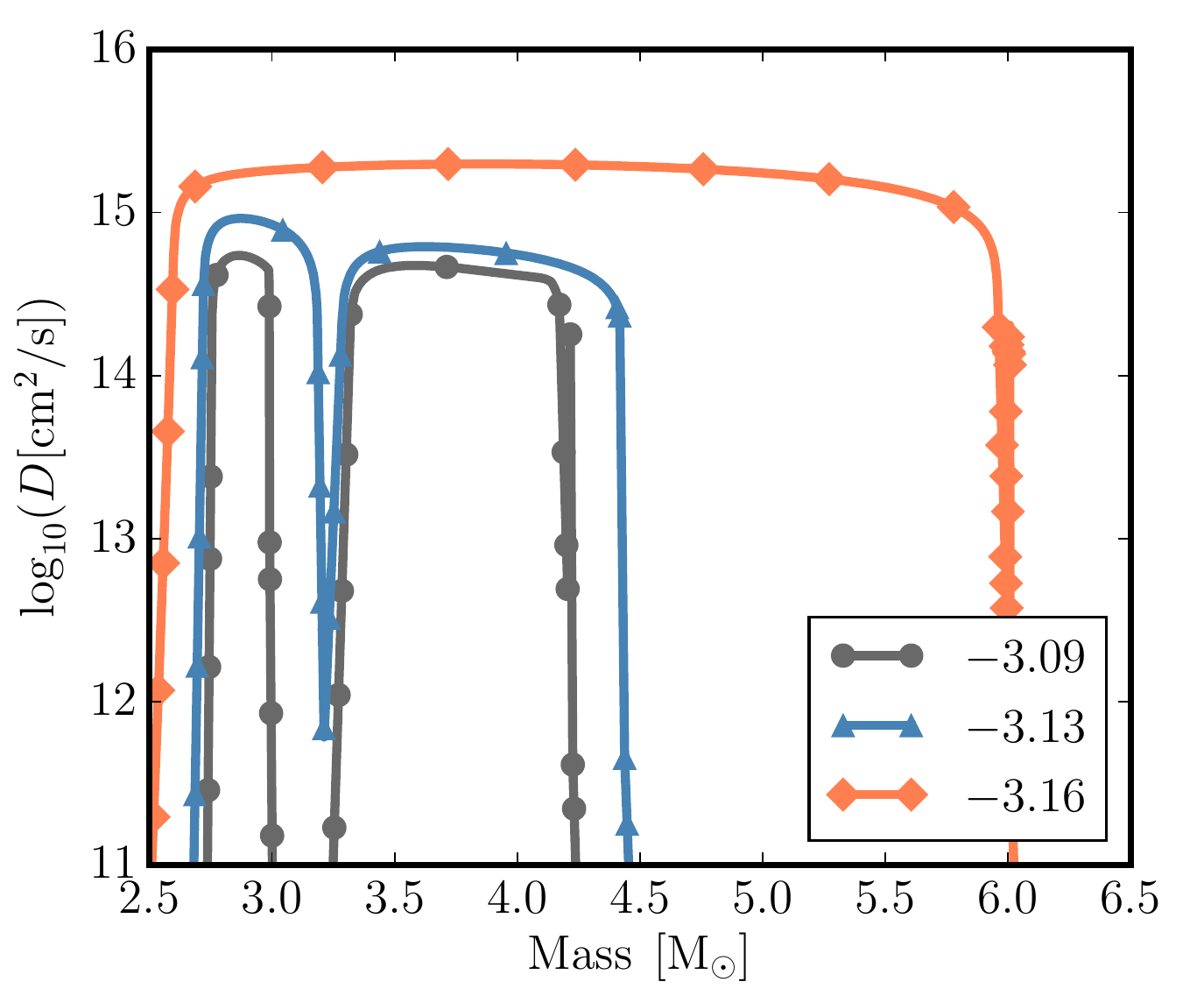}
	\caption{
		Log of the diffusion coefficient used in \mesa~over the Ne-O shell 
		merger found in the \Cc\ simulation. The values in the legend are 
		log of the time left until collapse, $\mathrm{log}_{10}(\tau - t)$, and 
		are taken at times over the merger. See 
		Figure~\ref{fig:3.1.2_kip_abu_C32} and Figure~\ref{fig:3.1.2_kip}.
	}
	\label{fig:3.1.2_logD} 
\end{figure}

\begin{figure}
	\includegraphics[width=\linewidth, clip=true, trim=0mm 3mm 0mm 4mm]{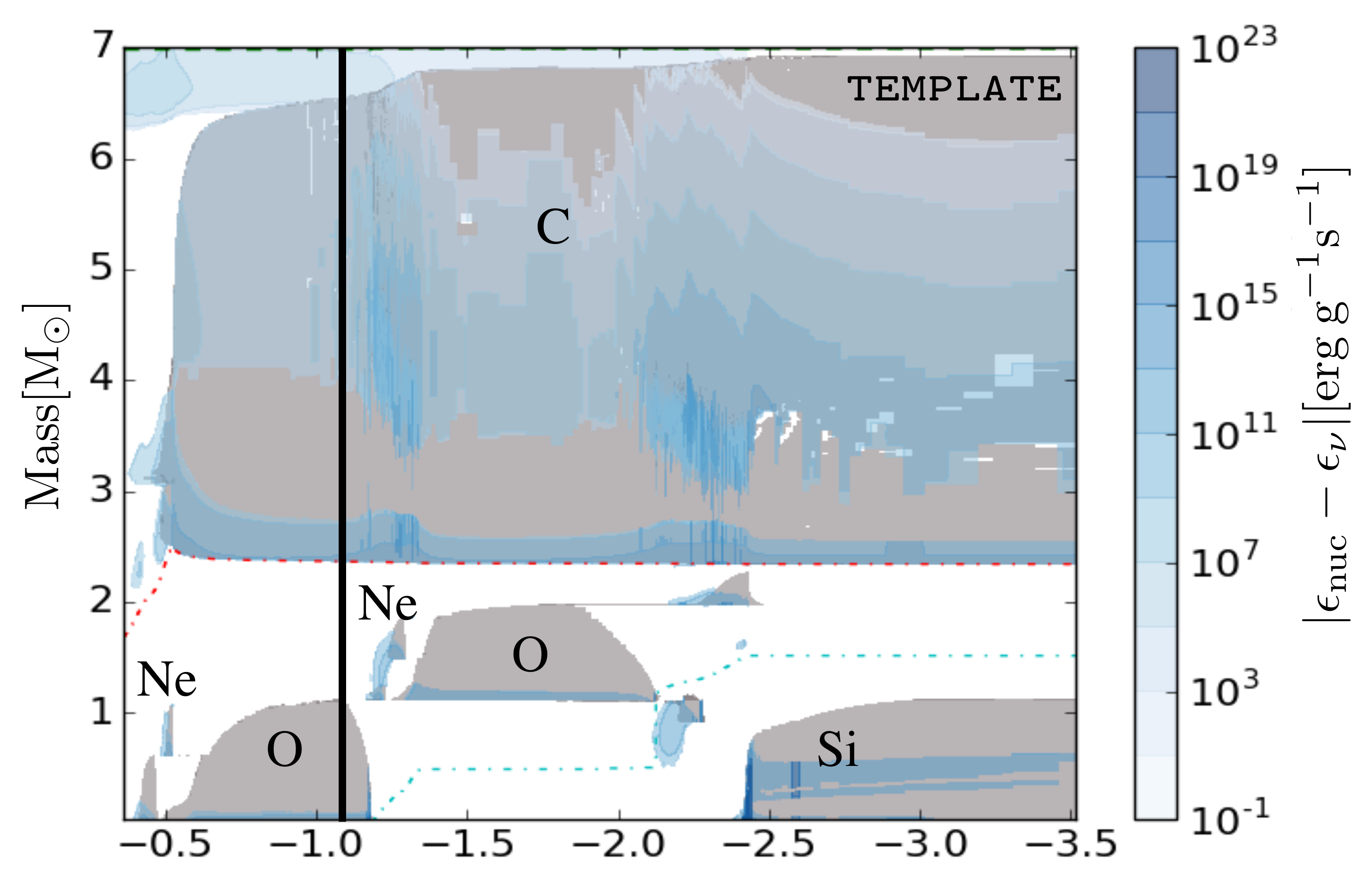}
	\includegraphics[width=\linewidth, clip=true, trim=0mm 3mm 0mm 4mm]{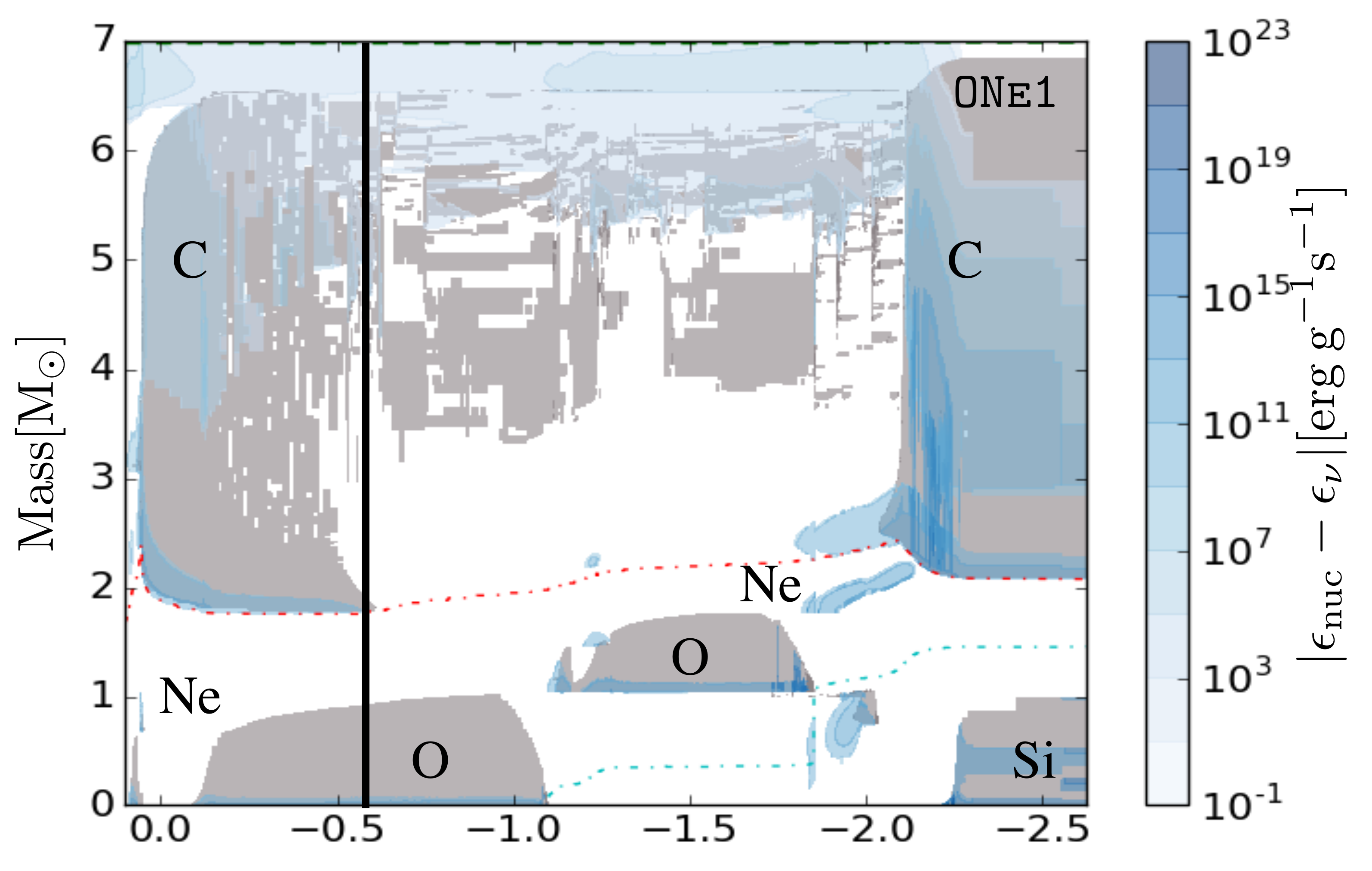}
	\includegraphics[width=\linewidth, clip=true, trim=0mm 3mm 0mm 4mm]{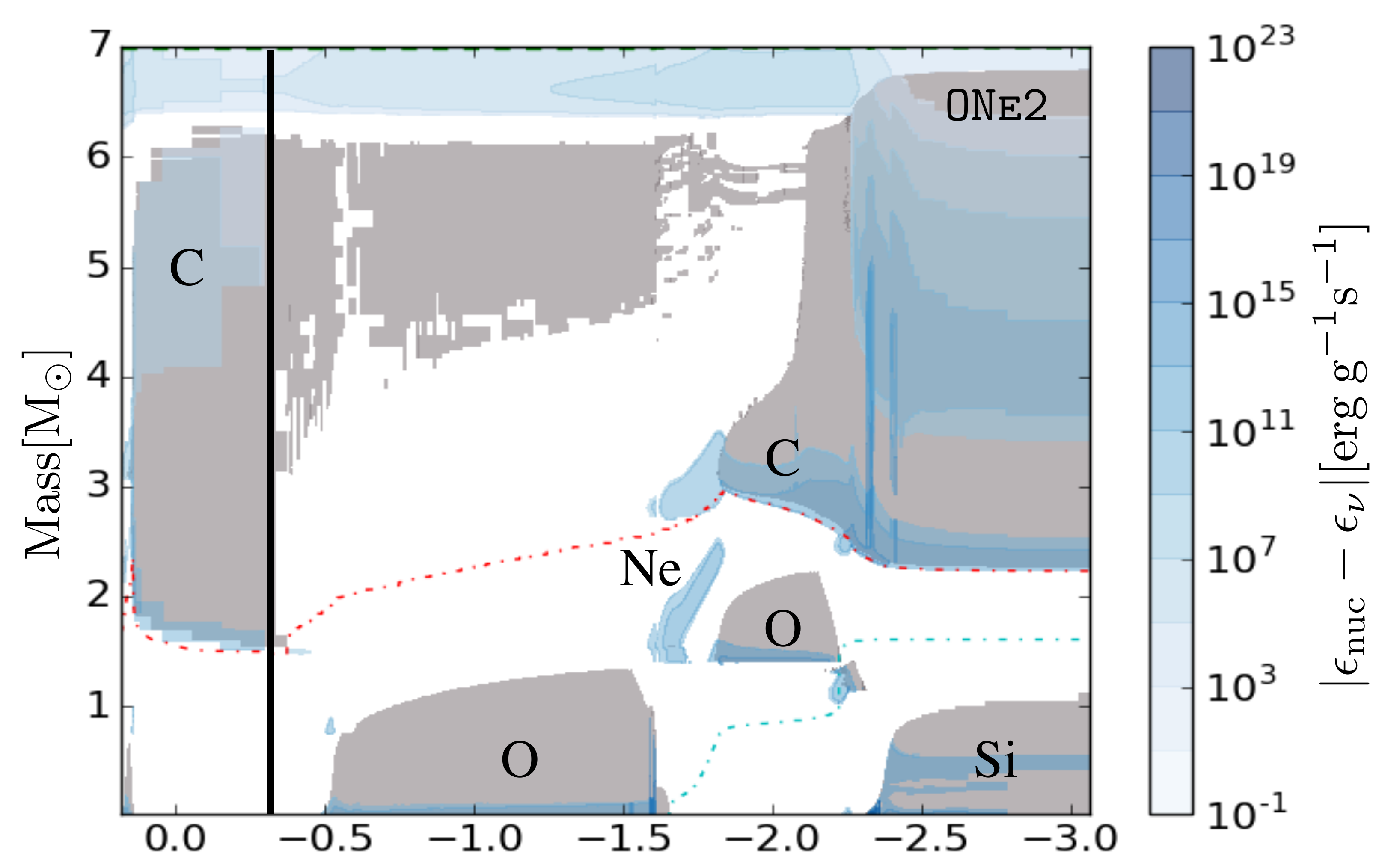}
	\includegraphics[width=\linewidth, clip=true, trim=0mm 2mm 0mm 4mm]{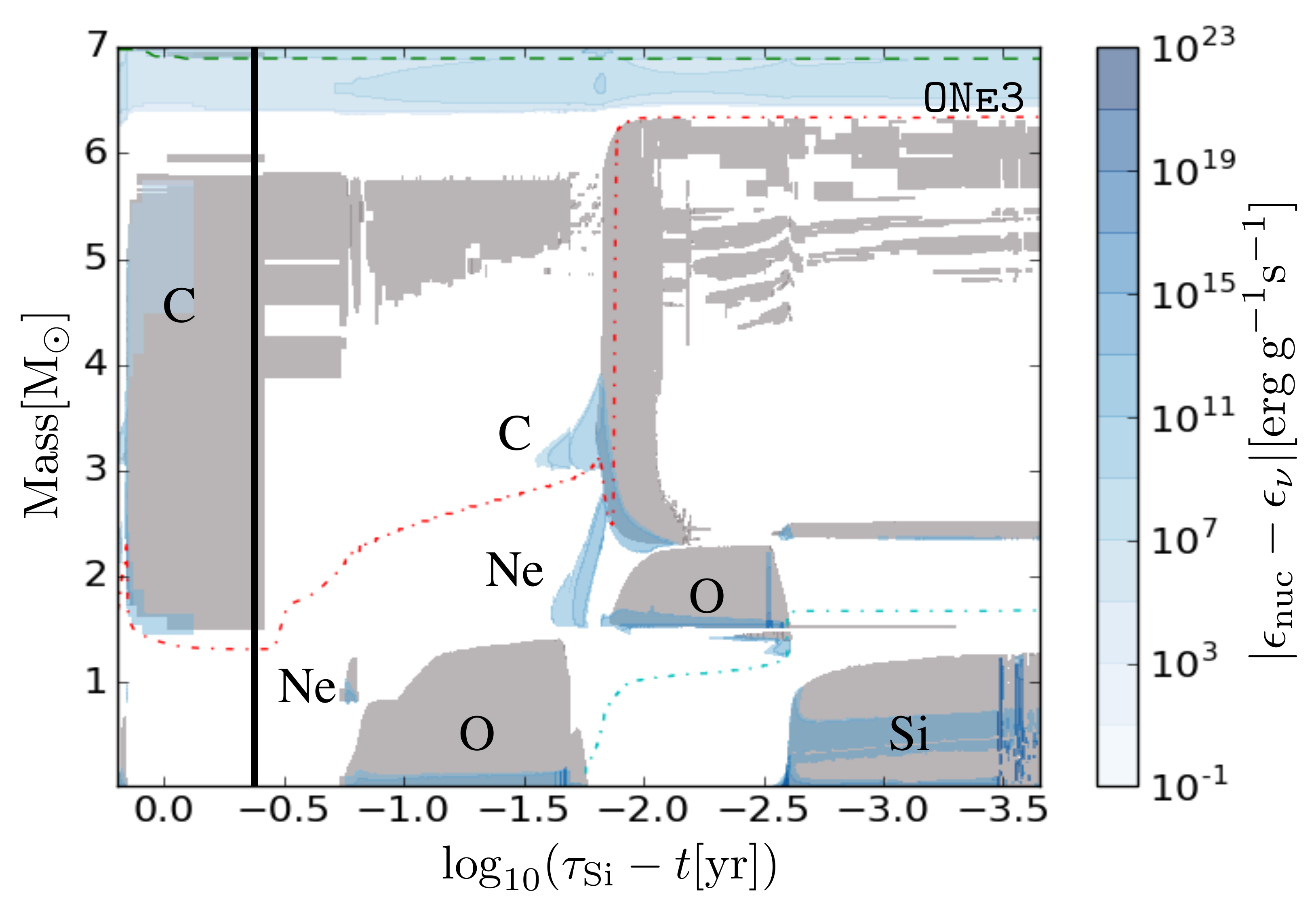}
	\caption{
		Kippenhahn diagrams for the \ONe\ simulations with increasing values 
		of \fcbm. The x-axis of the Kippenhahn 
		diagrams is the log of the time left until the end of convective Si core 
		burning. The solid line in the Kippenhahn diagram is the point where 
		the C shell that forms after Ne core burning reaches is maximum 
		depth in mass.
	}
	\label{fig:3.2.1_kip} 
\end{figure}

\begin{figure}
	\includegraphics[width=\linewidth, clip=true, trim=0mm 0mm 0mm 0mm]{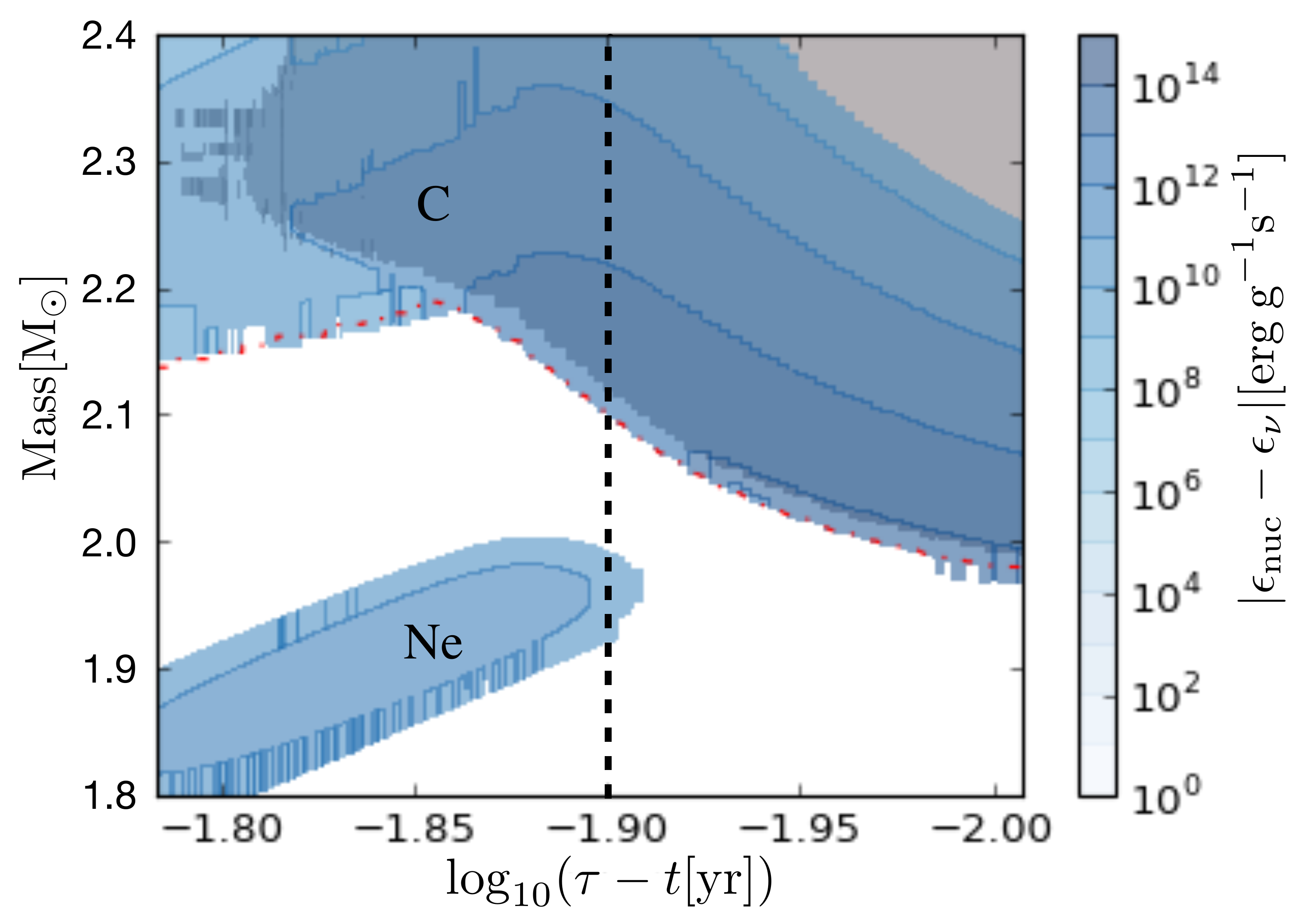}
	\includegraphics[width=\linewidth, clip=true, trim=0mm 0mm 0mm 0mm]{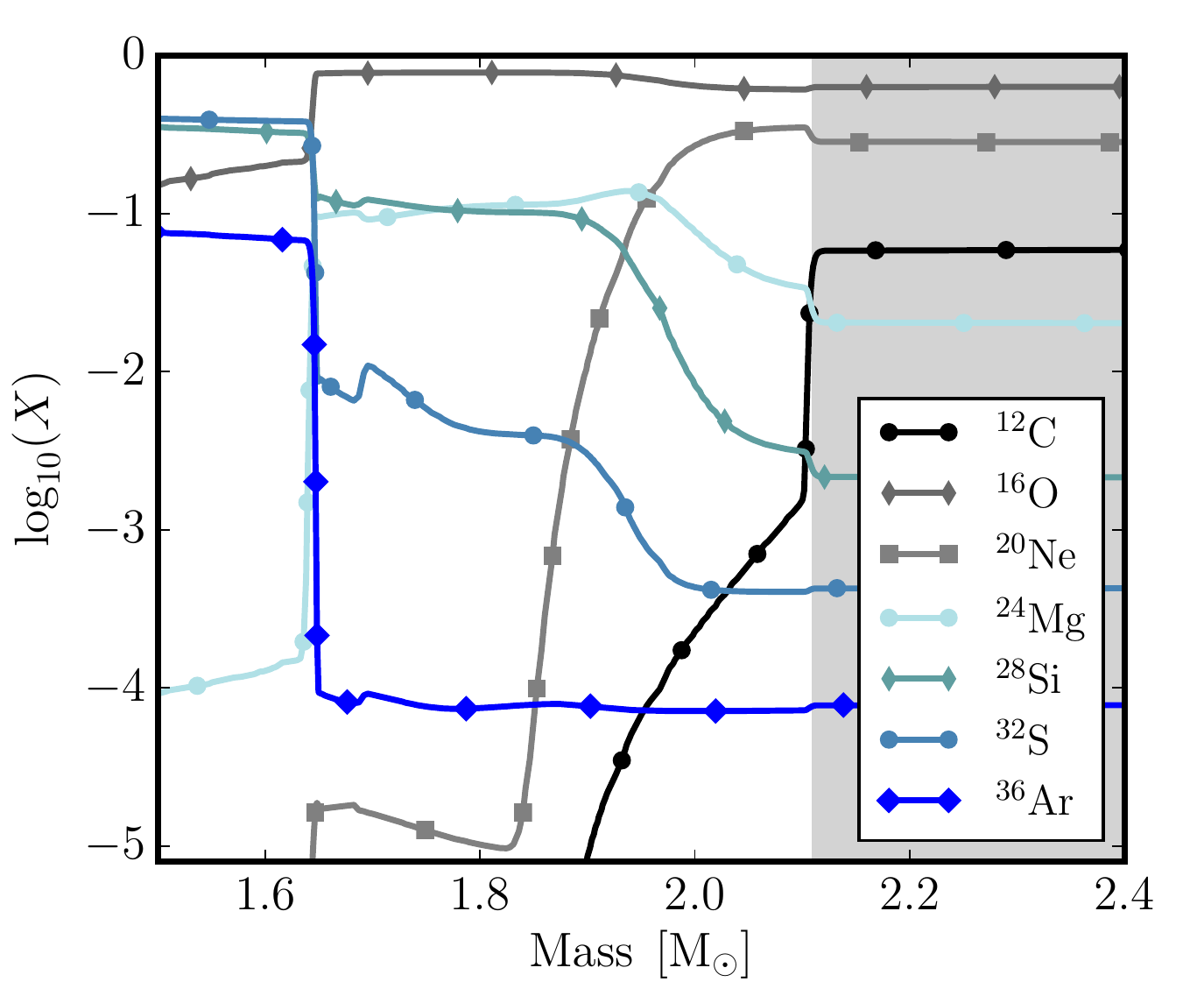}
	\caption{
		Kippenhahn diagram and abundance profile from the \Ca\ simulation. 
		The bottom radiative region is burning Ne, the upper convection zone 
		is a convective C shell. The black dashed line is the point where the 
		abundance profile is taken and is the same black dashed line as in Figure~\ref{fig:3.1.2_kip}. The abundance profile shows the ash 
		left behind from the radiative Ne burning as the C convection zone 
		mixes material in from the bottom.
	}
	\label{fig:3.1.2_kip_abu_C12} 
\end{figure}

Changing the CBM durning the late stage evolution changes the structure of the
simulations significantly. The \C\ simulations have different numbers of
convective C shells which change the ONe core mass and have consequences 
for the later evolution. The \ONe\
simulations experience Ne core-C shell interactions that change the number of C
shells and change the Ne and O core evolution.  Enhanced CBM in both of these 
post He core evolutionary stages change the structure near collapse.

The \template\ simulation has two large convective C shells during its
evolution (Figure~\ref{fig:3.1.1_kip}, \ref{fig:3.1.2_kip}). This simulation's
second convective C shell experiences limited entrainment from below so that
the ONe core mass, after this shell has developed, remains roughly constant at
about $2.34\Msun$ (Figure~\ref{fig:3.1.2_kip}, Table~\ref{tab:preSN_pars}).  In
the \Ca\ simulation, which experiences three convective C shells throughout
its evolution, the second convective C shell starts before Ne core burning.
After the convective O core forms, the bottom convective boundary of the second
C shell moves further out in mass, decreasing the mass contained by the
convection zone (Figure~\ref{fig:3.1.2_kip}). During O shell burning, the ONe
core mass increases to a maximum value of about $2.1\Msun$ before convective Si
core burning begins and a third convective C shell forms. The bottom boundary
of this shell begins to entrain material and the ONe core mass drops to around
$1.9\Msun$ before the end of the star's life (Table~\ref{tab:preSN_pars}). For
the \Cb\ simulation, four convective C shells develope during the
evolution. O core burning between the second and third convective C shells
pushes the ONe core boundary up to $\approx 2.7 \Msun$
(Figure~\ref{fig:3.1.2_kip}, \ref{fig:3.1.2_kip_abu}). When the third
convective C shell forms, entrainment from the bottom is large enough to reach
the ash left behind by radiative Ne burning, in the form of \oxygen,
\magnesium\ and \silicon\ (Figure~\ref{fig:3.1.2_kip_abu}). This C shell
reaches a depth of $2.27\Msun$ at the black dotted line in
Figure~\ref{fig:3.1.2_kip_abu}.  Although the convective boundary only reaches
the top of the Ne ash deposit, because the convective C shell spans a large
portion of the star, the \oxygen, \magnesium\ and \silicon\ that is mixed
into the convection zone is brought up to the top, at around $6.5\Msun$. The
\Cc\ simulation also develops four C convection zones during its evolution
(Figure~\ref{fig:3.1.1_kip}, \ref{fig:3.1.2_kip}). Where the \Cb\ simulation
experiences something like a classical \textit{dredge-up} (convection entrains
ash from a previous burning stage), in the \Cc\ simulation the convective Ne
and C shells merge after convective Si core burning (near the dotted line in
Figure~\ref{fig:3.1.2_kip}, panel 4). During Si core burning, a radiative Ne
shell forms at about $2.6\Msun$. At this point in time, C starts burning
radiatively at the ONe core boundary, at $\approx3.4\Msun$.  The
C shell becomes convective and begins to erode the ONe core
(Figure~\ref{fig:3.1.2_kip_abu_C32}). After this, the Ne shell also starts to
convect, mixing material in from the top. The difference in entropy between the
Ne and C shell is relatively small in this case, at $\Delta s/\mathrm{N_A kerg}
\approx 0.2$ (Figure~\ref{fig:3.1.2_kip_s}). In the \template\ simulation, 
a Ne shell forms
under the second C shell with an entropy difference of $\Delta s/\mathrm{N_A
kerg} \approx 2$. The
convective boundaries of the two shells meet and form one convection zone
spanning $4.4\Msun$ (Figure~\ref{fig:3.1.2_logD}). Relative to the C shell in
the \template\ simulation, the abundances of \magnesium\ and \silicon\
increase as the ash from form Ne burning is mixed into the convective C shell.

The \ONea\ and \ONeb\ simulations both experience three convective C shells
throughout their evolution as compared to the \template\ which has two
(Figure~\ref{fig:3.1.1_kip}, \ref{fig:3.2.1_kip}). In many of the simulations
in this study (\template, \Ca, \Cc, \ONea, \ONeb, and \ONec), convective Ne core burning is followed by the formation and growth of a convective C shell. In these simulations, near the end of Ne core burning, convection in the C shell begins near the ONe core boundary, at the composition profile imprint left by the previous convective C shell. As the Ne core burns out, material is mixed through the top and bottom boundaries of the C shell, as can be seen in Figure~\ref{fig:3.1.1_kip} and \ref{fig:3.1.2_kip}. The amount of CBM at the C shell boundaries determines the growth of this shell, and in the \Cc, \ONeb\ and \ONec\ simulations, can delay the formation of the convective O core (Figure~\ref{fig:3.1.1_kip}, \ref{fig:3.2.1_kip}). In the \template\ simulation
this does not happen. The second convective C shell remains until collapse with
a bottom boundary at $\approx 2.35 \mathrm{M}_{\odot}$. The convective O core
follows Ne core burning with a relatively short delay of $0.04 \mathrm{yr}$ or
about $15$ days. The \ONea\ simulation is similar
in that the delay from convective Ne core burning to convective O core burning
is also relatively short, $0.07 \mathrm{yr}$ ($26$ days). Although, for this
simulation, the second convective C shell does not last to collapse, it is
extinguished before the end of convective O core burning at a depth of $\approx
1.77\Msun$ (Figure~\ref{fig:3.2.1_kip}). The \ONeb\ simulation
has a delay of $1.07 \mathrm{yr}$ before convective O core burning. In this case the CBM in the second C shell pushes
the bottom boundary to a depth of $1.5 \mathrm{M}_{\odot}$. After the bottom
boundary of this C shell recedes, O core convection starts
(Figure~\ref{fig:3.2.1_kip}). The \ONec\ simulation is similar to \ONeb\ with
a delay of $1.26 \mathrm{yr}$ before convective O core burning. The convective
C shell then recedes after reaching $\approx 1.55 \mathrm{M}_{\odot}$ and
convection in the O core begins (Figure~\ref{fig:3.2.1_kip}). The O core formation delay in the \Cc\ simulation is shorter than both the \ONeb\ and \ONec\ simulations at $\approx 0.45 \mathrm{yr}$ or about 164 days. The C shell reaches a depth of $1.47\Msun$ before the O core becomes convective.

Relatively small changes in the shape of each zone, as a result of the enhanced CBM, produce large changes in the structure overall. Even with a small difference in CBM strength, as in the \C\ simulations, dredge-ups and shell mergers change the size of convective shells and the elemental distribution within them. Enhanced CBM in the ONe core can create a delay between the core Ne and core O burning on the order of $\approx 1/10$ of the Ne nuclear burning time scale, causing differences in the C shell formation above. Changing the CBM changes the structure of the simulations significantly, producing simulations with different elemental distribution, convective structure and time evolution.

\subsection{Dredge-ups, Shell Mergers and Core Masses}\label{sec:Dups}

The two main mechanisms by which CBM changes the core structure are
\textit{dredge-ups} and \textit{shell mergers}. Either one of these events
alone do not change the structure significantly, although the many
interacting dredge-ups and shell mergers over the evolution will. One of the
main consequences of increased CBM in the advanced burning stags of massive stars is to alter the relative masses of the burning shells in the stellar core.

A dredge-up occurs when entrainment at the bottom convective boundary mixes ash from a previous burning stage into the convection zone as in Figure~\ref{fig:3.1.2_kip_abu_C12}. The Kippenhan diagram shows the third C shell of the \Ca\ simulation entraining C ash from below. Entrainment from below mixes Ne ash, from the previous Ne burning region, into the C shell. In this case, because the C shells that develop in the late stage evolution of the star are large, spanning a few solar masses, the Ne ash can be mixed much closer to the surface. A shell merger occurs when a radiative region between two convection zones is eroded by entrainment, providing an opportunity for the two shell boundaries to meet and form one convection zone. Figure~\ref{fig:3.1.2_kip_abu_C32} shows a C-Ne shell merger found in the \Cc\ simulation. This shell merger is between the third Ne shell, which forms at the end of the second O shell, and the fourth C shell. After the two shells merge, C can be transported to the bottom of the Ne burning region and Ne ash can be transported upward closer to the surface. During the evolution of the simulations, dredge-ups and shell mergers interact and cause the largest changes in the simulations. Although in some cases, one dredge up may not change the structure drastically, the cumulative effects of many interacting dredge-ups and shells mergers throughout the evolution create the large differences seen in these simulations.

\begin{figure}
	\includegraphics[width=.5\textwidth]{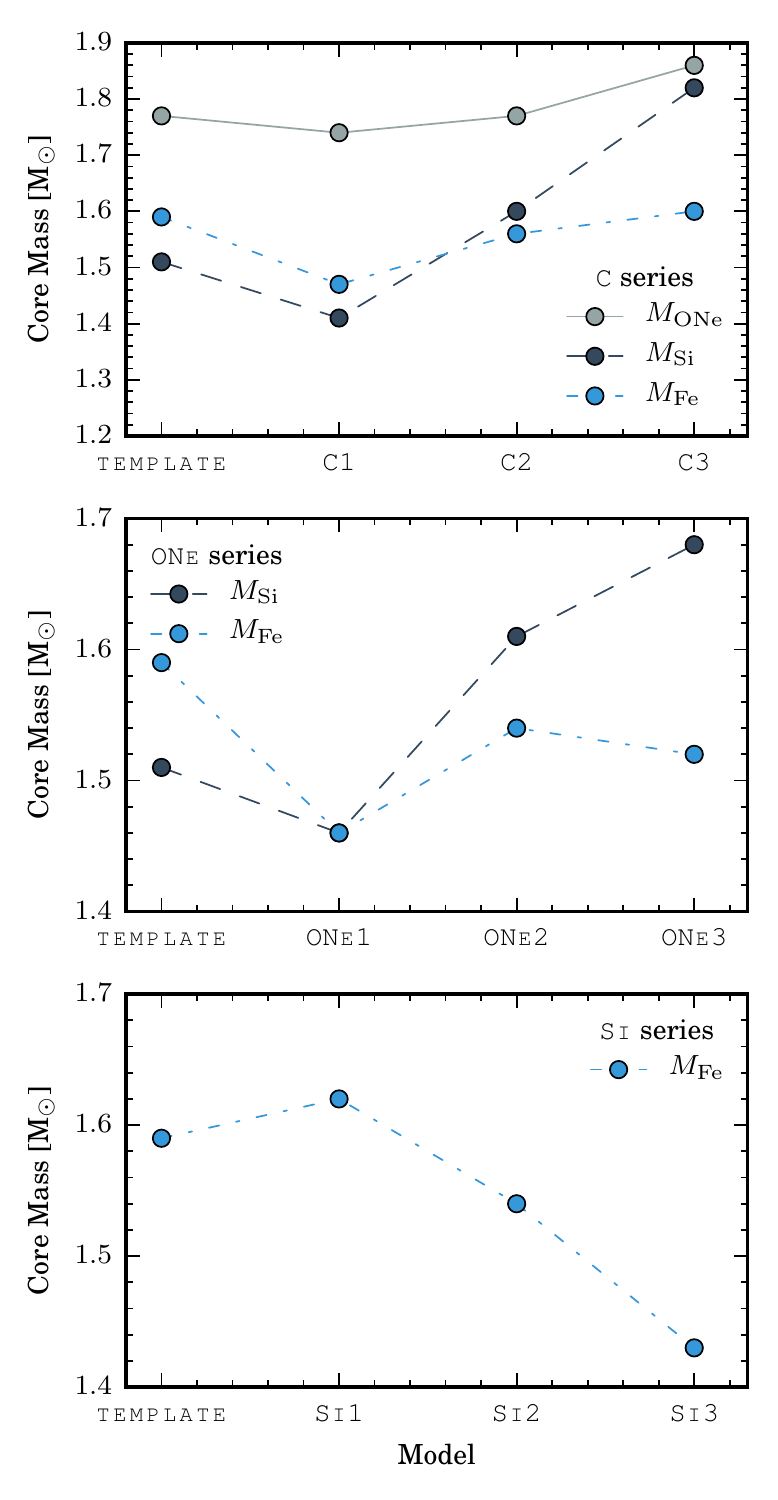}
	\caption{
		ONe, Si and Fe core masses at Ne-ignition, Si-depletion and
		about 30 seconds before collapse, respectively. Refer to 
		Table~\ref{tab:par} for more details of these values. All of the
		core masses show a non-linear dependence on the
		convective boundary mixing parameter \fcbm.
	}
	\label{fig: core-masses} 
\end{figure}

Table~\ref{tab:par} gives the CO, ONe, Si and Fe core masses of the 10 simulations, along with a summary of the \fcbm\ values used to compute them. The same information is represented graphically in Figure~\ref{fig: core-masses}. An example of the differences caused by the C shell dredge-ups can be seen in the \C\ simulations (Figure~\ref{fig: core-masses},  upper panel). In the \C\ simulations, once the first convective C shell forms, it grows as material from above and below is mixed into the convection zone. Entrainment from above mixes \carbon, \neon\ and \oxygen\ into the convection zone where the \carbon\ can burn. From below, entrainment erodes the ONe core formed by previous radiative C burning, mixing \neon, \magnesium\ and \magnesiumB\ into the convection zone. At the bottom boundary, the amount of entrainment increases for larger values of \fcbm\ deepening the C shell in mass coordinate. This increased entrainment can reduce the ONe core mass experienced by the next convective C shell, and subsequently, the ONe core mass at Ne ignition in the \template\ and \Ca\ simulations. In the \Cb\ and \Cc\ cases, a second C shell develops before the onset of core Ne burning, which increases the ONe core mass at Ne ignition (Figure~\ref{fig:3.1.1_kip}). Although the \C\ simulations have a difference in ONe core masses of $\Delta M_{\mathrm{ONe}} = 0.12 \mathrm{M}_{\odot}$, the cumulative effects of increased CBM create a $\Delta M_{\mathrm{Si}} = 0.41 \mathrm{M}_{\odot}$.

Even though the effect of one dredge-up or shell merger produces relatively small changes to the simulations in most cases, the interaction of many of these events throughout the evolution can produce simulations that are unrelatable in the convective structure, having different convection zones at different mass coordinates.

\subsection{Nucleosynthesis} \label{sec: profac}

\begin{figure}
	\includegraphics[width=0.5\textwidth]{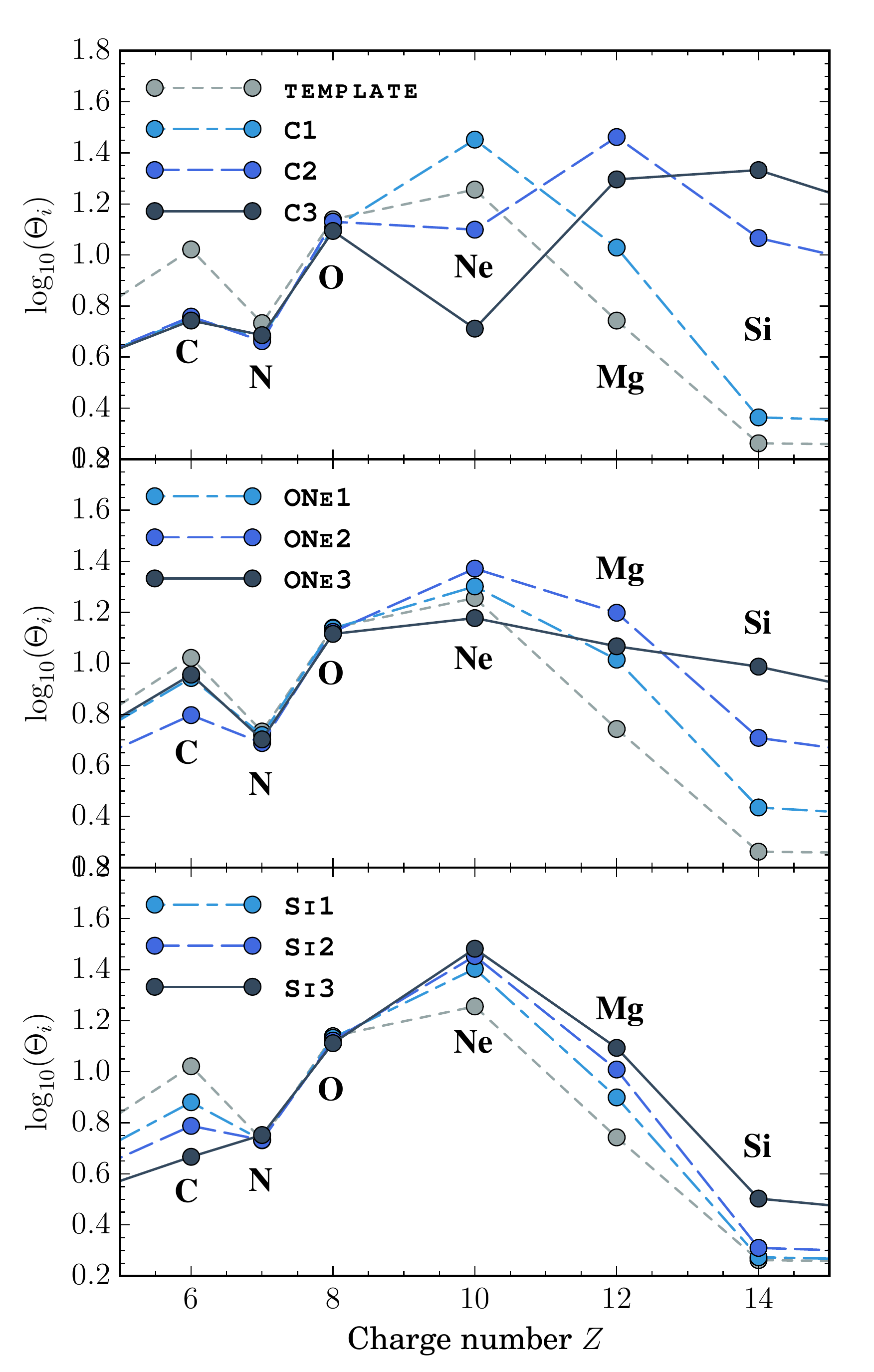}
	\caption{
		Log of the presupernova overproduction factors for
		the \C, \ONe\ and \Si\ simulations. The panels contain the simulations 
		from the run set as well as the \template~simulation.
	}
	\label{fig:profac} 
\end{figure}

\begin{figure}
	\includegraphics[width=\linewidth, clip=true, trim=0mm 3mm 0mm 3mm]{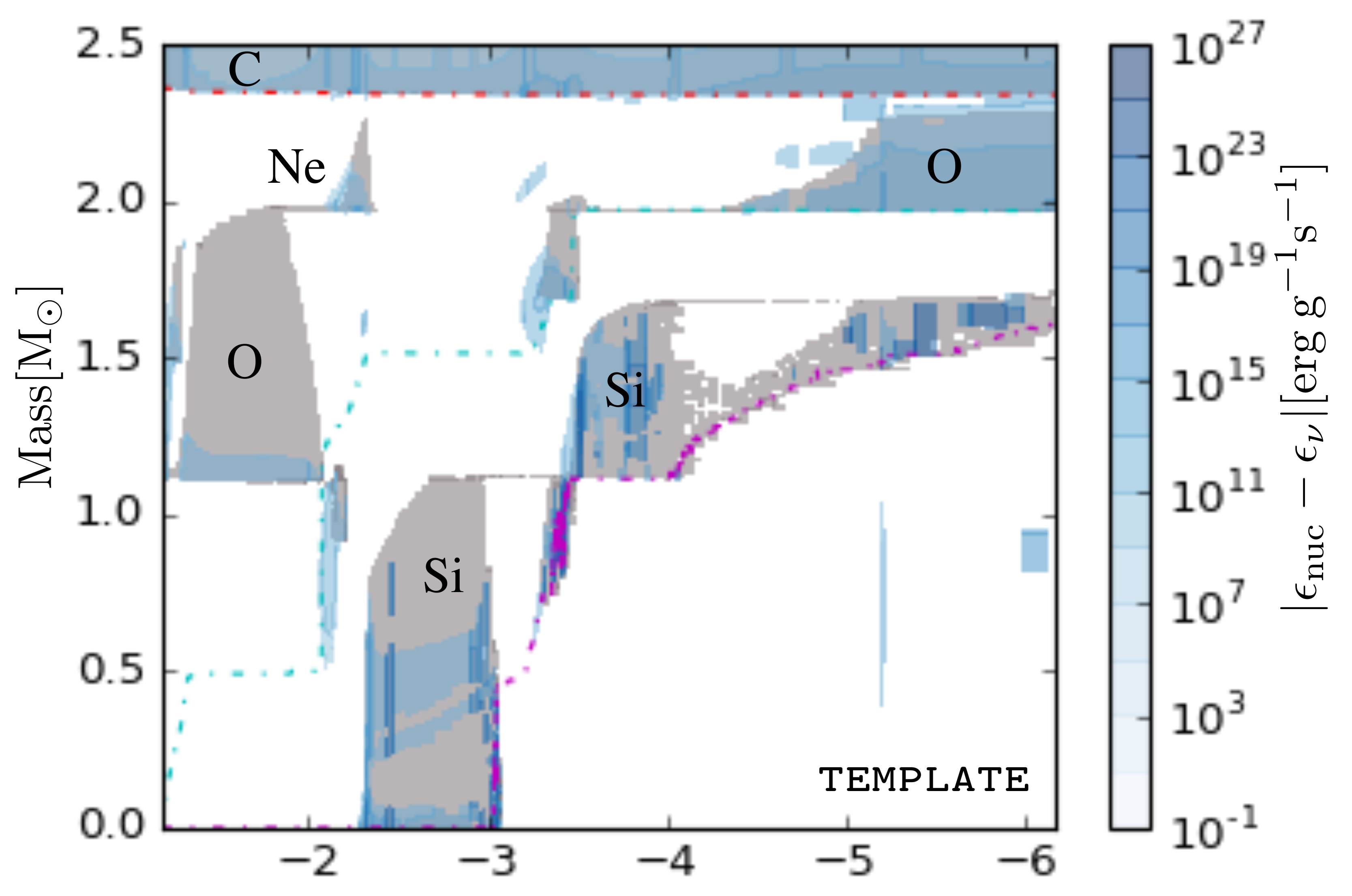}
	\includegraphics[width=\linewidth, clip=true, trim=0mm 3mm 0mm 3mm]{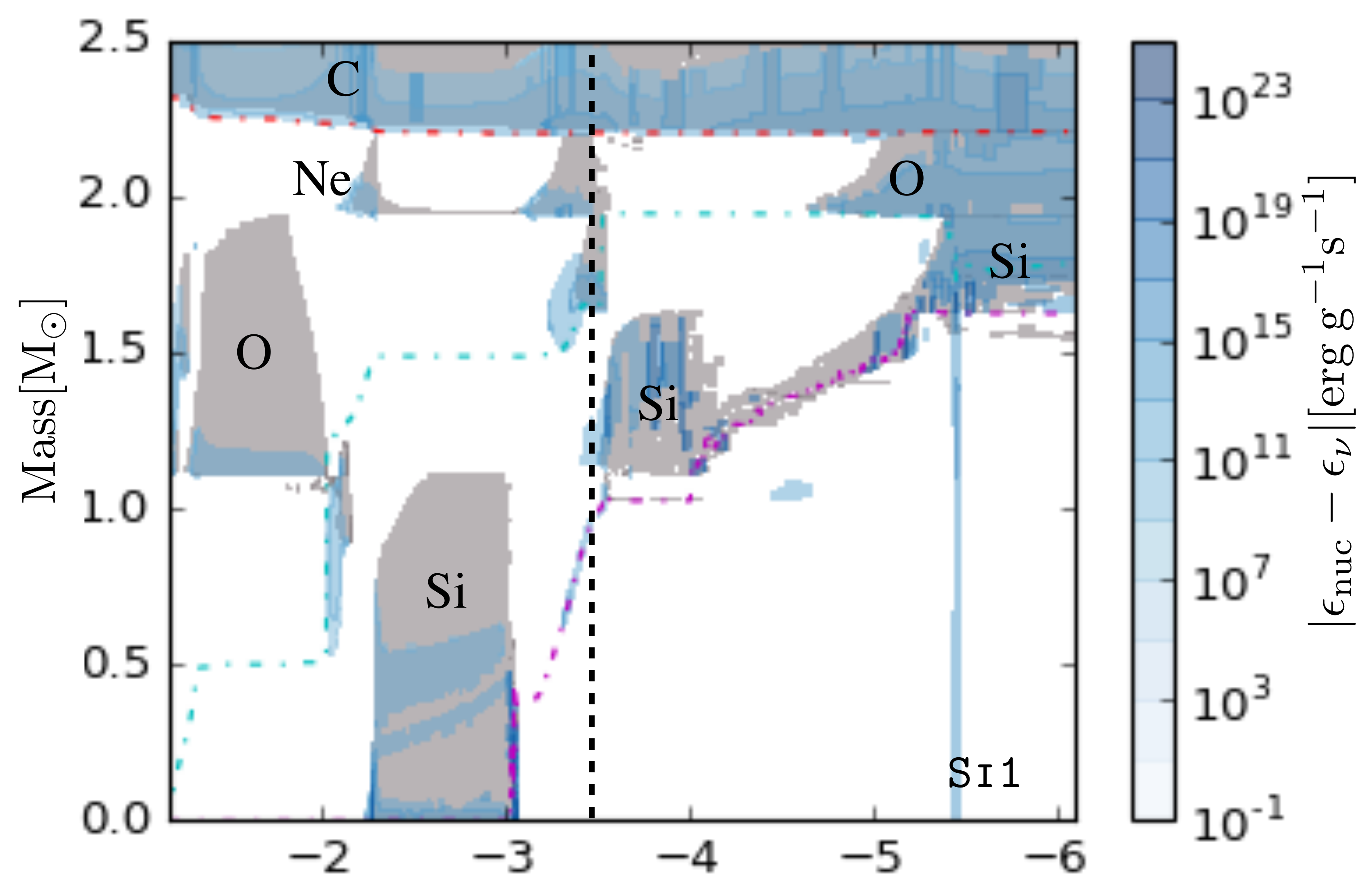}
	\includegraphics[width=\linewidth, clip=true, trim=0mm 3mm 0mm 3mm]{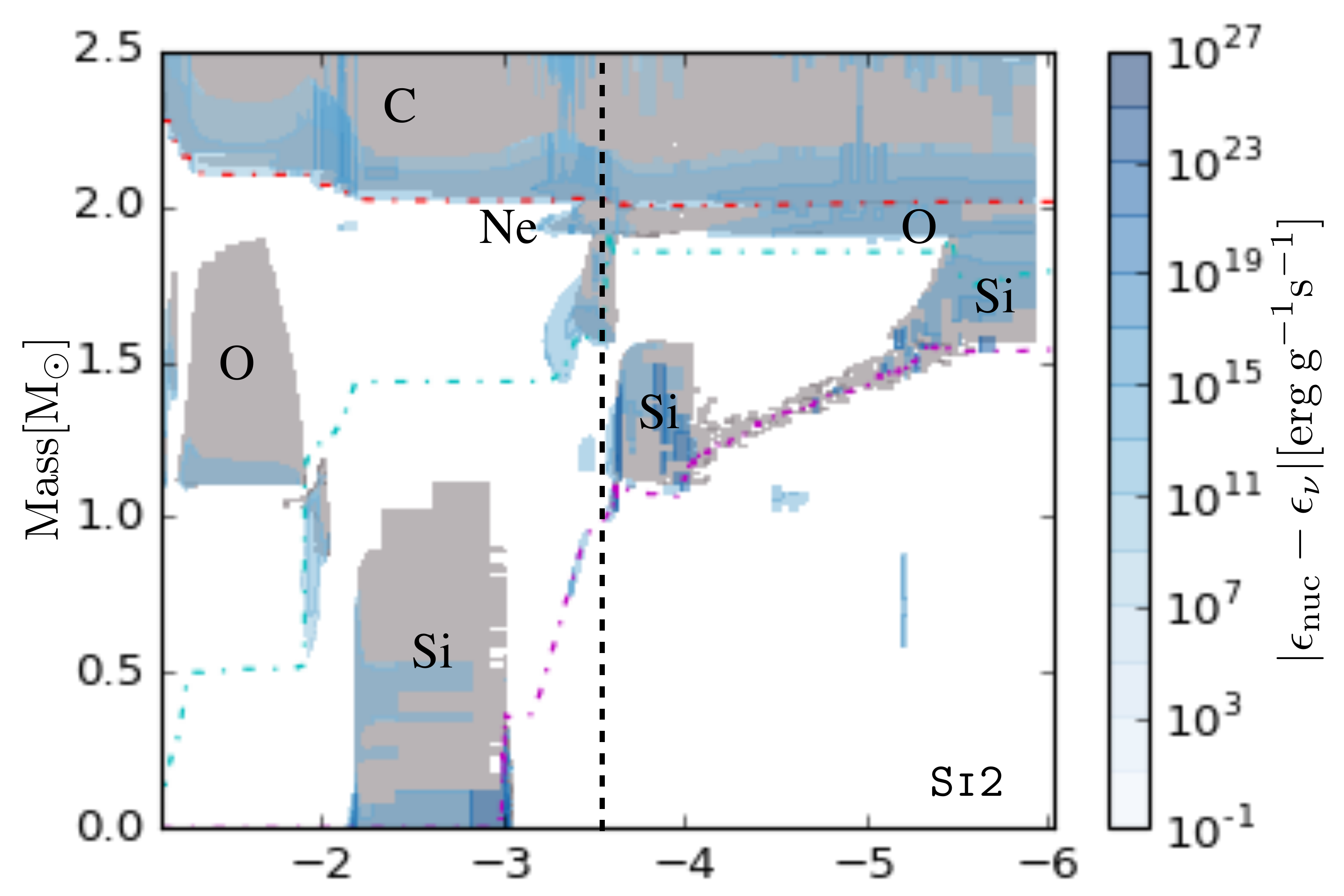}
	\includegraphics[width=\linewidth, clip=true, trim=0mm 2mm 0mm 3mm]{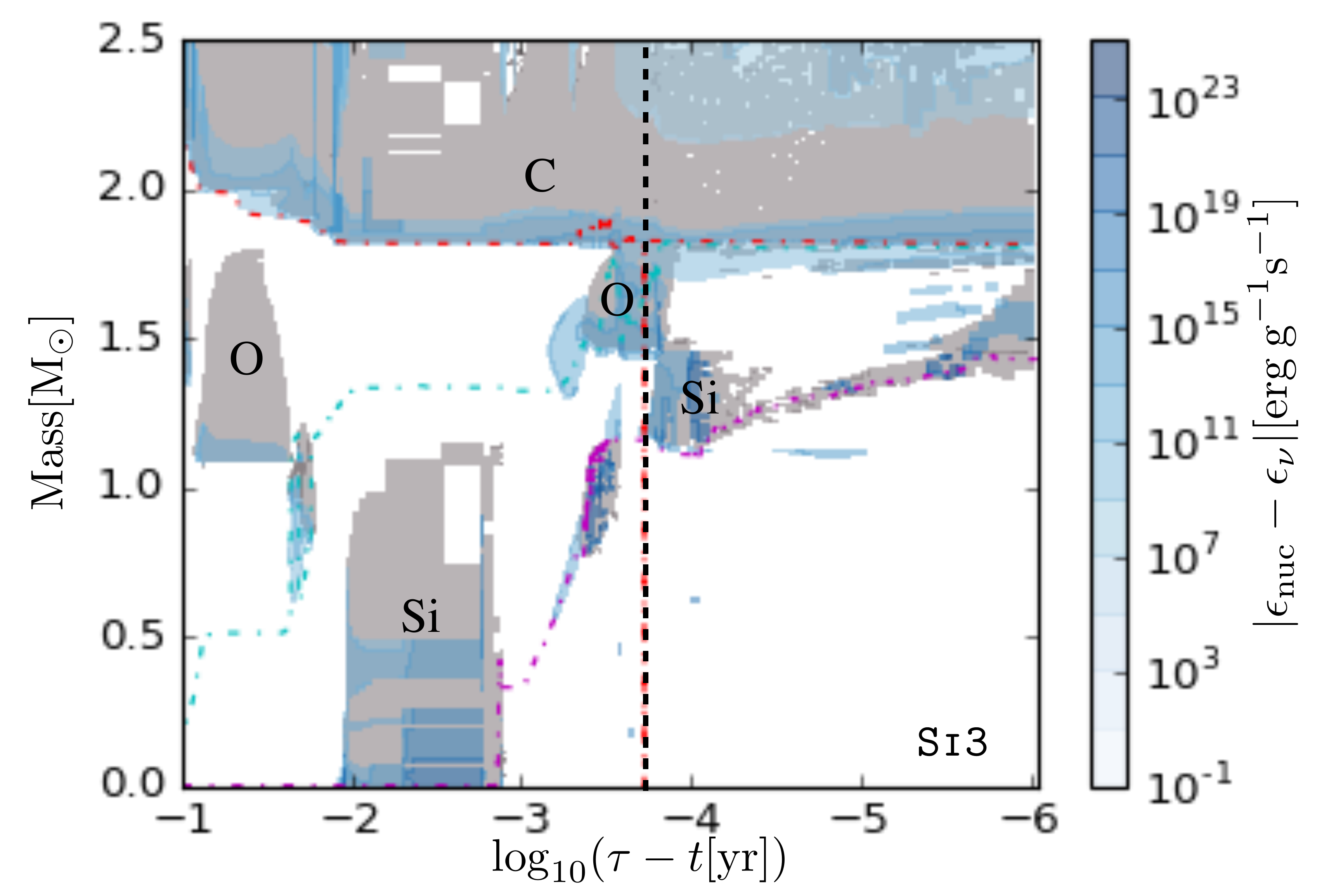}
	\caption{
		Kippenhahn diagrams for the \Si\ simulations. Starting from the top, 
		the diagrams are for the \template, \Sia, \Sib\ and \Sic\ simulations. 
		The black dashed lines are where the entropy profiles are taken in 
		Figure~\ref{fig:3.3.1_s}.
	}
	\label{fig:3.3.1} 
\end{figure}

\begin{figure}
	\includegraphics[width=\linewidth, clip=true, trim=0mm 0mm 0mm 0mm]{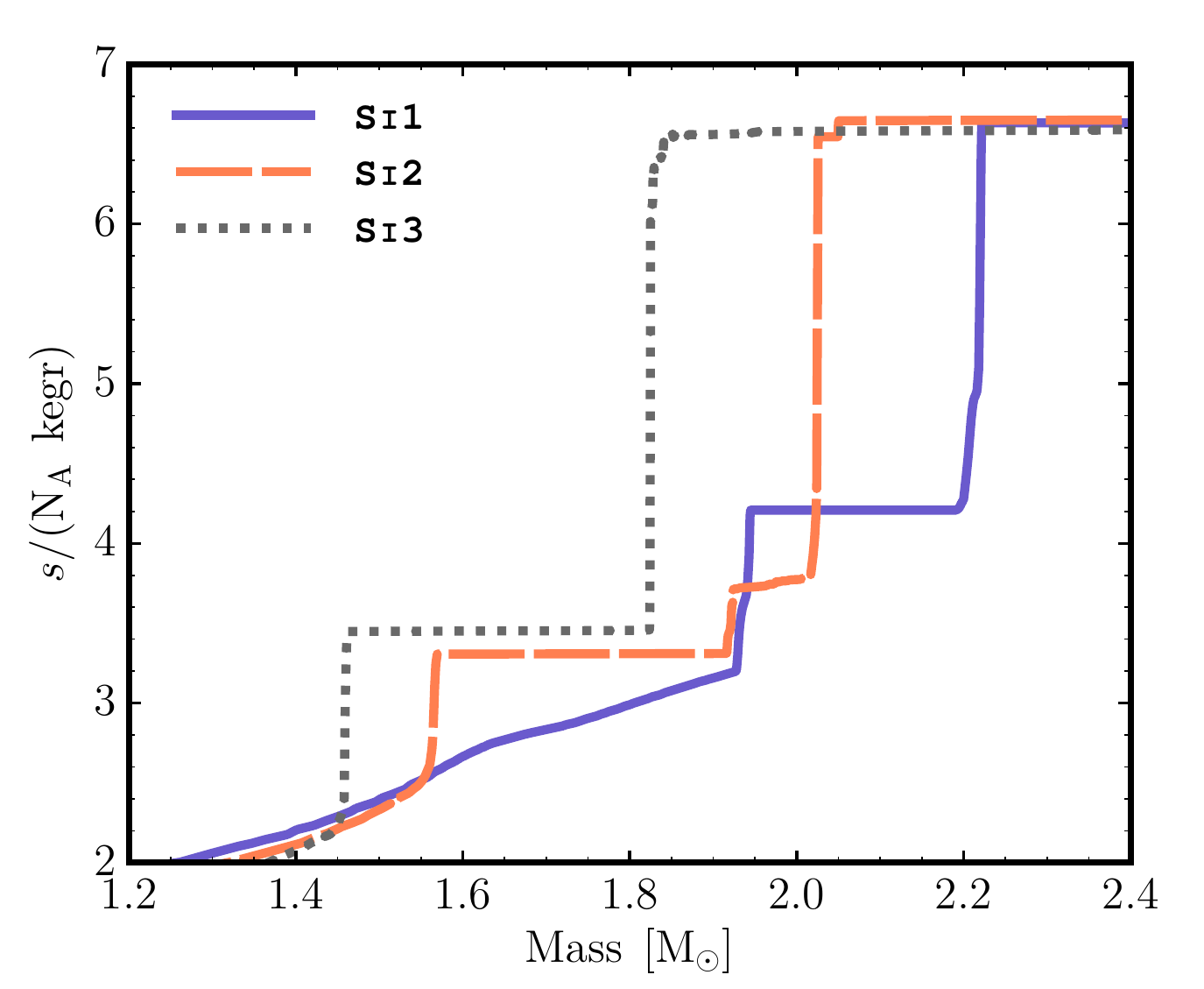}
	\caption{
		Entropy for the \Sia, \Sib\ and \Sic\ simulations for the bottom of 
		the C shell. The profiles are taken at the black dashed lines in 
		Figure~\ref{fig:3.3.1}.
	}
	\label{fig:3.3.1_s} 
\end{figure}

\begin{figure}
	\includegraphics[width=\linewidth, clip=true, trim=0mm 0mm 0mm 0mm]{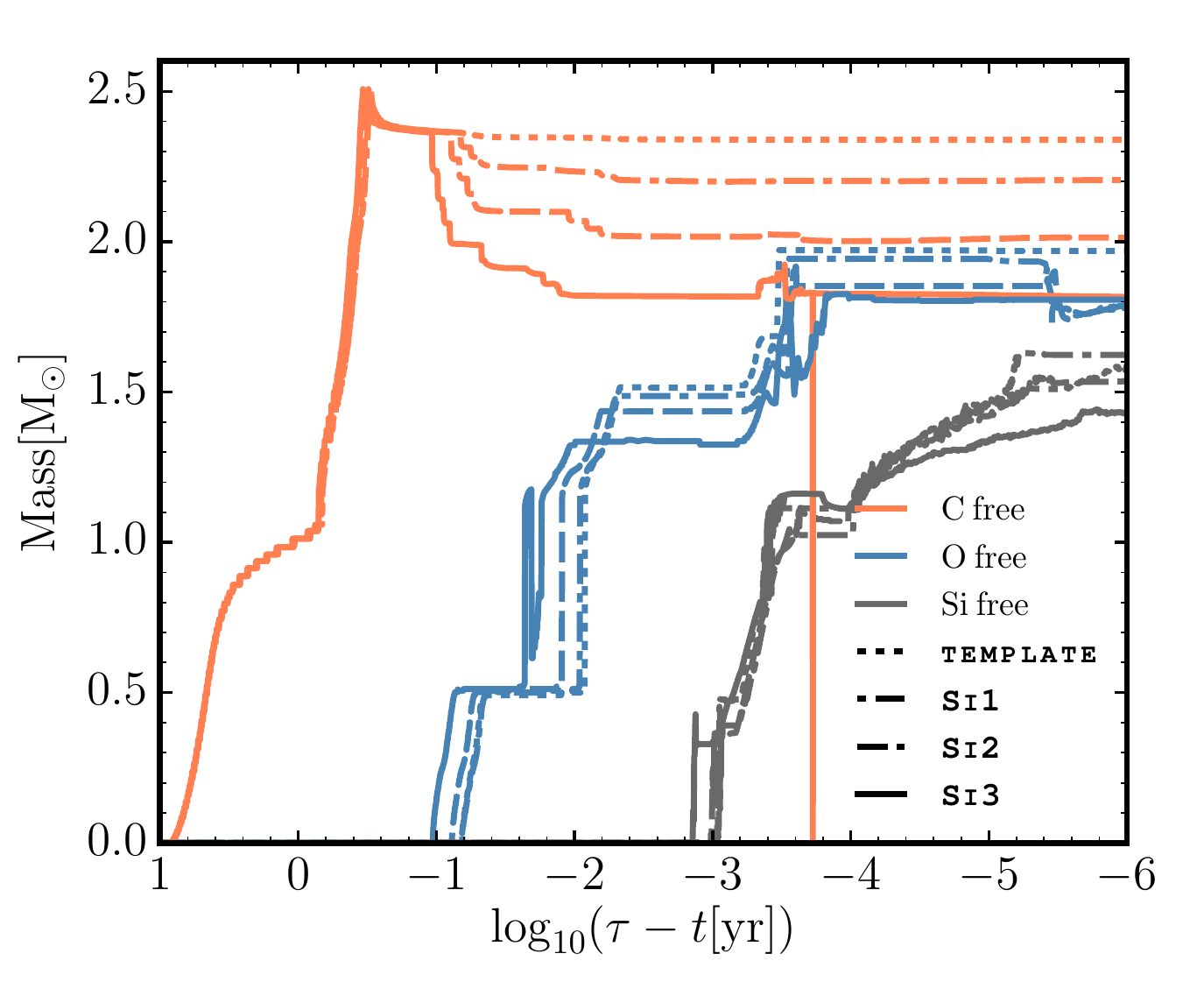}
	\caption{
		The ONe, Si and Fe core masses for the \Si\ simulations. The 
		x-axis is log of the time left until collapse and the y-axis is the 
		enclosed mass from the core. The plot shows the differences 
		in the core masses with respect to the amount of CBM implemented 
		during core Si burning. The small differences in the lines before core 
		Si burning is due to differences in the evolutionary time for each simulation.
	}
	\label{fig:3.3.1_cm} 
\end{figure}

During a Type II supernova explosion, the supernova remnant in the form of a
black hole or neutron star, will be composed of material that previously made up
the core of the star. This material consists of some of the metals produced in 
the core, and will not be ejected by the supernova explosion
\citep[see, e.g.][and references therein]{Colgate1966,Fryer2009,Janka2012,
Ugliano2012,Ertl2016}. This means that a significant amount of the material 
the star produces will not be ejected and used for further star formation. 

To estimate the effects of CBM on the chemical yields from massive stars during
the advanced burning stages, a mass cut is used. The mass cut is an estimate of 
the Lagrangian coordinate
separating the compact remnant from the ejected material. Here, the mass cut is
taken from the formula of \citet{Fryer2012} for the \emph{delayed} explosion
scenario. This formula gives a remnant mass of 5.7\Msun~for the 25\Msun,
solar-metallicity model considered in this work. Had the \emph{rapid} explosion
scenario been considered, the entire star would fall back into the remnant, and
only the material from the envelope that has been ejected by winds would enrich
subsequent star formation. The value of 5.7\Msun~puts the mass cut near the top
of the C shell for all simulations. Because the C shell spans a large range of
masses, and is convective in most simulations, small variations to this mass
cut (on the order of a solar mass or more) would not significantly change the
chemical signatures found in the ejected material. The work by \citet{Sukhbold2016,Farmer2016}, and the work here show that the final state of stellar simulations is dependent on more than just ZAMS mass and metallicity. Therefore, determining the mass cut in this way is an approximation and is meant to illustrate how the nucleosynthetic signatures closer to the surface are affected by the CBM in the later stages of evolution (see Section~\ref{sec: summary}).

In order to investigate the net effects of CBM on the nucleosynthesis of each simulation, the ejected mass of each element is determined from the stellar evolution model by the presupernova overproduction factors, $\Theta_{i}$. These overproduction factors \citep{Pignatari2016,Ritter2017} are given by
\begin{equation}
\Theta_{i} = \frac{1}{M_{\mathrm{sol}, i}}{\int^{M^{\prime}_{\mathrm{surf}}}_{M_{\mathrm{delay}}}Z^{\prime}_{i}(m)dm},
\end{equation} \label{eq:2}
where the primes denote values taken at collapse, and $Z_{i}$ is the abundance of element $i$, $M_{\mathrm{delay}}$ is the mass cut, $M^{\prime}_{\mathrm{surf}}$ is the mass of the simulation at collapse and $M_{\mathrm{sol}, i}$ is the mass of the element from the initial abundance, given by
\begin{equation}
M_{\mathrm{sol}, i} =\int^{M^{\prime}_{\mathrm{surf}}}_{M_{\mathrm{delay}}}Z_{i}(m)dm = Z_{i}(M^{\prime}_{\mathrm{surf}} - M_{\mathrm{delay}}),
\end{equation} \label{eq:3}
for a uniform initial composition. The values for $\Theta_{i}$ give the mass of the element $i$ above the mass cut normalized to the initial composition of the model. Note that the $\Theta_{i}$ values have not been processed by the supernova shock or include any contributions from the winds. Figure~\ref{fig:profac} gives
$\Theta_{i}$ for the \C, \ONe~and \Si~simulation sets. The greatest
deviation from the \template\ simulation is found in the \C\ simulations. In
the \C\ set, the amount of C does not significantly change between the 
\C~simulations, where as the \template\ simulation is enhanced by a factor of
10. Ne, Mg and Si also show large nonlinear variation due to the complexity of
the Ne and C shell interactions (Section~\ref{sec: Convective structure} and
\ref{sec:Dups}).  The \Cc\ simulation produces the least amount of Ne where as
the \Ca\ simulation produces the most, with a variation of less than one order of
magnitude. Simulations with the lowest Ne production (\Cb\ and \Cc) produce
the most Mg and Si where simulations with high Ne (\template\ and \Ca) produce
lower amounts of Mg and Si. This implies that in the \Cb\ and \Cc\ cases the
abundances in the C shell shows evidence of Ne burning from below. For the
\ONe\ simulation set, the \template, \ONea\ and \ONeb\ simulations all show
similar trends in $\Theta_{i}$. C decreases for enhanced \fcbm\ and Ne, Mg
and Si all increase, in part due to the depth of the third C shell
(Section~\ref{sec: Convective structure}). The \ONec\ simulation does not
follow this trend. The $\Theta_{\mathrm{Ne}}$ value is less than that for the
\template, and Mg and Si are both higher. Si in particular is larger than that from the \template\ with a value of $\Theta_{\mathrm{Si}} = 9.7$, compared to $\Theta_{\mathrm{Si}} = 1.8$ for the \template. The $\Theta_{i}$ for the
\Si\ simulations have the least deviation from the \template.
Figure~\ref{fig:profac} shows that $\Theta_{\mathrm{C}}$ decreases for
increasing values of \fcbm, where as Ne and Ne ash both increase. Unlike the
\ONe\ simulations, the \Si\ simulations don't show large dredge-ups of Ne ash
into the C shell (Figure~\ref{fig:3.2.1_kip}, \ref{fig:3.3.1}). In the \Sia,
\Sib\ and \Sic\ simulations, only a limited amount of mixing between the C
shell and the underlying Ne shell is possible due to the large entropy
gradients found between the shells (see Section~\ref{sec:3D hydro},
Figure~\ref{fig:3.3.1_s}). In these simulations, $\Theta_{\mathrm{C}}$ shows the
largest variation from that of the \template. During core Si burning, Ne and O
shells form under the C shell and promote C shell dredge-ups of C ash
(Figure~\ref{fig:3.3.1}). This decreases the ONe core mass and the depth of the
C shell boundary (Figure~\ref{fig:3.3.1_cm}), increasing the temperature at the
bottom of the C shell.

The CBM during the later stage evolution of these simulations can affect the
$\Theta_{i}$ values by mixing material processed in metal burning shells into the
C shell. Simulations that show large deviations from the \template\ are those
that have dredge-ups and shell mergers with the C shell, and have enough time
left in their evolution to mix that material passed the mass cut.

\subsection{Presupernova core masses and compactness parameter} \label{sec: compactness}

\begin{figure}
	\includegraphics[width=.5\textwidth]{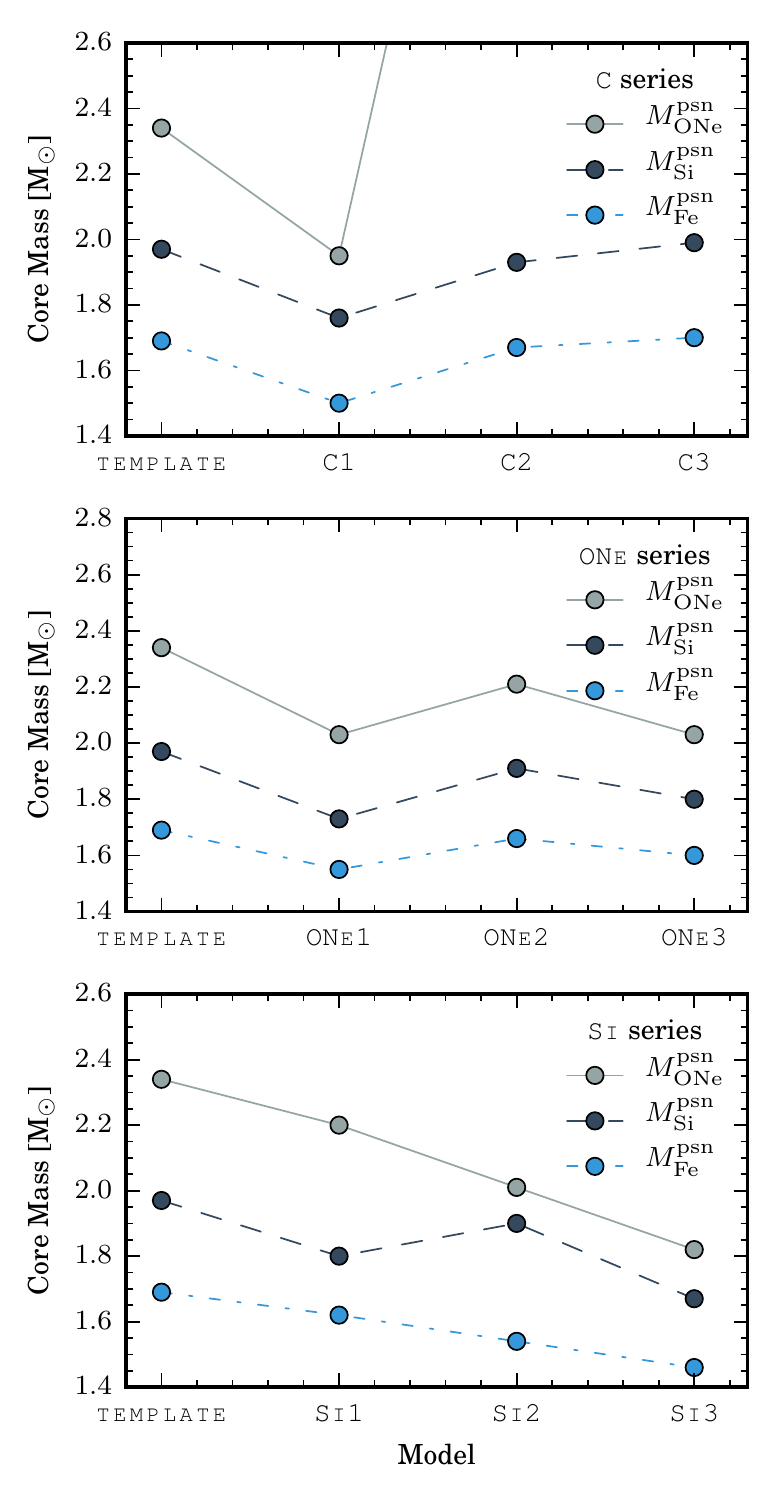}
	\caption{
		ONe, Si and Fe core masses at the presupernova stage for 
		each simulations. The values 
		are also given in Table~\ref{tab:preSN_pars}. 
		The ONe, Si and Fe core masses show a non-linear 
		dependence on the convective boundary mixing parameter, 
		\fcbm,~in the \C~and \ONe~simulations. This non-linearity is due 
		to the accumulation of structural changes after a number of shell-burning 
		episodes have passed. The Fe core mass does show a linear trend with 
		\fcbm~in the \Si~series, due to the limited amount of time left for 
		convective shells to interact before collapse. The ONe core mass is 
		extremely large in the \C~series owing to the transport of O and 
		Ne to the edge of the CO core during the merging of the convective 
		Ne and C shells.
	}
	\label{fig: presn-core-masses} 
\end{figure}

\begin{table}
	\centering
	\caption{Presupernova core masses and compactness parameter
		$\xi_{2.5}$. The core masses are taken when the infall velocity
		in the Fe core reaches $1000 \mathrm{km/s}$. The core masses
		are defined in terms of lower limits of key abundances, these
		definitions can be found in the caption of
		Figure~\ref{fig:2_kip_runs}. The presupernova compactness,
		$\xi_{2.5}$, taken at $\mathrm{log}_{10}(\tau - t) = -6$, the 
		values do not change significantly past this point. The last two 
		rows give the range and percent difference (\% diff.) of the smallest and 
		largest values compared to the the \template~simulation. The 
		range is defined to be the absolute difference between the 
		highest and lowest values.
	}
	\begin{tabular}{lcccc}
		\toprule
		Name & 
		$M^{\mathrm{psn}}_{\mathrm{ONe}} [\Msun]$ &
		$M^{\mathrm{psn}}_{\mathrm{Si}} [\Msun]$  &
		$M^{\mathrm{psn}}_{\mathrm{Fe}} [\Msun]$  &
		$\xi_{2.5}$
		\tabularnewline
		\midrule
		\template & 2.34 & 1.97 & 1.69  &  0.272  \\ 
		\Ca       & 1.95 & 1.76 & 1.50  &  0.172  \\ 
		\Cb       & 4.36 & 1.93 & 1.67  &  0.304  \\ 
		\Cc       & 3.95 & 1.99 & 1.70  &  0.354  \\ 
		\ONea     & 2.03 & 1.73 & 1.55  &  0.159  \\ 
		\ONeb     & 2.21 & 1.91 & 1.66  &  0.249  \\ 
		\ONec     & 2.03 & 1.80 & 1.60  &  0.152  \\ 
		\Sia      & 2.20 & 1.80 & 1.62  &  0.217  \\ 
		\Sib      & 2.01 & 1.90 & 1.54  &  0.162  \\ 
		\Sic      & 1.82 & 1.67 & 1.46  &  0.120  \\ 
		\midrule
		range [\Msun]    & 2.54 & 0.32 & 0.24  &  0.234  \\
		\% diff & (22.2, 86.3) & (15.2, 0.01) & (13.6, 0.06) &  \\
		\bottomrule
	\end{tabular}
	\label{tab:preSN_pars}
\end{table}

\begin{figure*}
	\includegraphics[width=\linewidth]{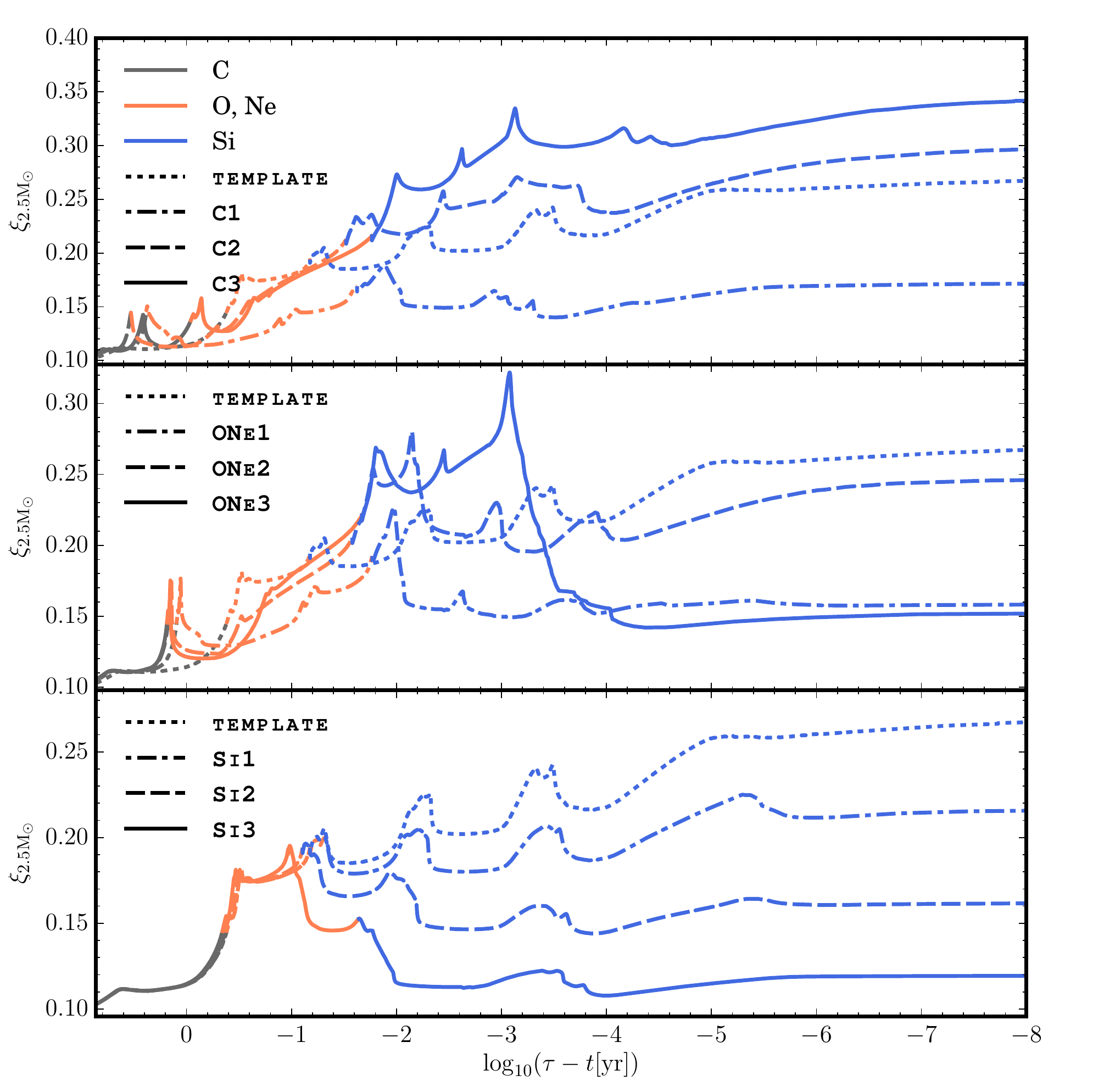}
	\caption{
		Evolution of the compactness parameter for the three sets of simulations. The x-axis
		in given in log of the time left until collapse. The colours of the
		line represents the core burning stage where grey is C core burning,
		orange is Ne and O core burning and blue is Si core burning. Each set
		of simulations diverges from the template soon after the enhanced CBM
		is implemented. Differences in the evolutionary compactness before this
		are due to differences in the age of the stars being simulated.
	}
	\label{fig:compact} 
\end{figure*}

The core masses of the models at the presupernova stage are shown in
Figure~\ref{fig: presn-core-masses} and given in Table~\ref{tab:preSN_pars}.
The \C\ simulations show a large increase in the ONe core mass for the \Cb\ and 
\Cc\ simulations (Figure~\ref{fig:3.1.2_kip}). This ONe core mass increase corresponds to the large entrainment of 
Ne burning ash into the C shell. In the \Cb\ simulation, this happens during the C shell
dredge up that takes place after radiative Ne shell burning (see 
Section~\ref{sec: Convective structure}, Figure~\ref{fig:3.1.2_kip_abu}). In the \Cc\ simulation, this increase happens 
after the Ne-C shell merger (see Section~\ref{sec: Convective structure}, \ref{sec:Dups}, Figure~\ref{fig:3.1.2_kip_abu_C32}). The Si and Fe core masses 
of the \C\ and \ONe\ simulations are non-monotonic with increasing \fcbm. This is 
due to the cumulative interaction of the C, Ne and O shells that each simulations 
experiences, changing the core structure of the star significantly before collapse 
(Section~\ref{sec:Dups}). In the \Si\ simulations, due to the lack of time left in the evolution compared to the convective turn over time scales of shells, and the presence of a well established C shell, the convective shells do not interact as in the \C\ and \ONe\ simulations. Rather than being the result of many dredge-ups and shell mergers, the changes in the \Si\ simulation core masses are dominated by the C shell dredge-ups that happen at the beginning of Si core burning (Figure~\ref{fig:3.3.1}). These dredge-ups are larger with increasing \fcbm\ as can be seen in the presupernova ONe core masses (Table~\ref{tab:preSN_pars}).  In these simulations, both the presupernova ONe core masses and the Fe core masses are monotonically decreasing with increasing \fcbm. Similar effects of CBM on the core of the simulations can be seen in the compactness of the massive star core.

The compactness parameter of a massive star core is a measure of the depth of
the gravitational potential well at the bounce phase of a core-collapse
supernova. A correlation has been found between this parameter and whether 
or not the stellar model will produce a supernova explosion when exploded using 
one dimensional codes
\citep{OConnor2011,Ugliano2012,Ertl2016,Muller2016,Sukhbold2016, Sukhbold2017}. 
Lower values of the compactness favour explosions while higher
compactness favour weak or failed explosions that likely result in black hole
formation \citep{Ugliano2012,Ertl2016}. \citet{OConnor2011} define the bounce 
compactness of a stellar core as
\begin{equation}
\label{eq:2}
\xi_M = \frac{M/\mathrm{M}_{\sun}}{R(M)/1000 \mathrm{km}} \biggr\vert_{t=t_{\mathrm{bounce}}}
\end{equation} 
where $M$ is the baryonic mass and $R(M)$ is the radius in which a mass of 
$M$ is enclosed at the time that the infalling core material bounces off the 
proto-neutron star ($t_{\mathrm{bounce}}$). To evaluate the compactness, the 
mass is generally set to $M = 2.5\Msun$, i.e.  the relevant mass scale for black 
hole formation \citep{OConnor2011}. Alternatively, evaluating the compactness at 
the presupernova stage, when the infall velocity of the iron core reaches 
1000kms$^{-1}$, provides a reasonable estimate for the value of 
$t_{\mathrm{bounce}}$. \citet{Sukhbold2014} found that, among other things, the 
non-monotonicity of the presupernova compactness with respect to the 
progenitor's ZAMS mass is strongly dependent on the behaviour of the C and O 
burning shells (more recently \citet{Sukhbold2017}). In this study it is found 
that the position and timing of these shells are, in turn, dependent on the strength 
of the CBM used to compute them. 

It is important to note that the bounce compactness given by \citet{OConnor2011} is not a definitive determination of the explodability of massive star simulations. Other effects such as asymmetries, turbulence and even the dimension in which the explosion is calculated all affect the explodability \citep{Dolence2013,Radice2017,Muller2016,Ertl2016}. Never the less, calculating the value of $\xi_{2.5}$ allows for comparison to the work of \citet{Sukhbold2017} and \citet{Farmer2016}.

Investigating how the compactness changes throughout the evolution of the star
with respect to CBM provides more insight into the non-monotonicity found by
\citet{Sukhbold2014}. The compactness throughout the core evolution, the 
\textit{evolutionary compactness}, is plotted in Figure~\ref{fig:compact} and 
the quantity $\xi_{2.5}$ at the presupernova stage is given in 
Table~\ref{tab:preSN_pars}. Each stellar lifetime plotted in 
Figure~\ref{fig:compact} shows spikes in the evolutionary compactness where 
the value increases rapidly, followed by dips where the value decreases. The 
spikes are caused by contractions within $2.5\Msun$ as a convective burning 
event ends. This decreases the radius at which $2.5\Msun$ is encloses, and 
increases the compactness. At the end of a core contraction, a burning phase 
begins, whether that be a shell or the core. The convective regions that result 
from the burning expand in radius pushing $2.5\Msun$ further out and 
decreasing the compactness value. 

Changing the CBM strength changes the value of $\xi_{2.5}$ significantly. For 
the simulations branched at C burning, the value of compactness ranges from 
0.17 to 0.32. The \Ca\ simulation, with the smallest values of \fcbm, has the 
largest deviation from the presupernova compactness of the \template\ 
simulation, with $\xi_{2.5}=0.17$ for \Ca, and $\xi_{2.5}=0.27$ for the \template. 
Both the \Cb\ and \Cc\ simulations have increasing presupernova compactness 
with increasing \fcbm\ ($\xi_{2.5}$ of 0.30 and 0.35 receptively). In the \Ca\ 
simulation, at the beginning of convective Si core burning, the C shell experiences 
a dredge-up that drops the C shell boundary to $\approx 1.95\Msun$ 
(Figure~\ref{fig:3.1.2_kip}, Table~\ref{tab:preSN_pars}). This C shell boundary 
corresponds to the bottom of the convective C shell during this time. The 
bottoms of similar convective shells in the \template, \Cb\ and \Cc\ simulations 
are located at $\approx 2.34\Msun$, $2.17\Msun$ and $2.25\Msun$ receptively. 
In the \Cb\ and \Cc\ simulations, these convective shells are burning C and Ne 
although they serve the same purpose in expanding the material contained 
within them. The depth of the convective C shell dominates the compactness 
in the \Ca\ simulation as a significant amount of the material enclosed within 
$2.5\Msun$ is in this convection zone, decreasing the compactness value. 
Although in the \template\ simulation the C shells bottom boundary has the 
largest mass, convective O and Si shell develop during the later stages of the 
evolution, expanding the core and decreasing the compactness. 

The \ONe~simulations show more non-monotinicity than that in the \C~simulations.
The values of $\xi_{2.5}$ range from 0.15 to 0.27 with the \template\
simulation being the largest and the \ONec\ simulation being the smallest. In
this case, the \ONea\ and \ONec\ simulations have similar values, with
$\xi_{2.5}=0.16$ for the \ONea\ simulation, while the \ONeb\ simulation has a
value closer to the \template\ at $\xi_{2.5}=0.25$. Examining the bottom 
convective boundary mass location of C burning (or C, Ne and O burning) 
convection zones near $2.5\Msun$ shows that simulations with similar 
presupernova compactness values have similar ONe core mass values. 
In the \ONe\ simulation set, at the end of the evolution, the ONe core mass 
corresponds to the bottom boundary mass of these large convection zones. 
The \template\ and \ONeb\ simulations have a bottom convective boundary 
masses of $2.34\Msun$ and $2.21\Msun$ receptively (with $\xi_{2.5}$ of 0.27 
and 0.25). The \ONea\ and \ONec\ simulations both have bottom convective 
boundary masses of $2.03\Msun$ (with $\xi_{2.5}$ of 0.16 and 0.15). In this 
case, similar to the \C\ simulations, the more material that is contained in the 
large convection zone around $2.5\Msun$, the lower the value of $\xi_{2.5}$ 
will be due to the expansion of material in that convection zone. The effect of 
expansion and contraction of the core on the compactness can be seen in the
evolutionary compactness plot for the \ONec\ simulation. This simulation 
experiences a sharp spike in compactness around 
$\mathrm{log}_{10}(\tau-t)\approx -3$ ($\mathrm{log}_{10}(\tau_{\mathrm{Si}}-t) 
\approx -1.6$ in Figure~\ref{fig:3.2.1_kip}). Leading up to this spike in the 
compactness evolution, the only significant energy generation within the 
inner $6\Msun$ of the CO core comes from convective Si core burning 
which extends to $\approx 1.3\Msun$. Above the convective Si core, 
the material is contracting, increasing the evolutionary compactness. The 
Si core burning ends abruptly and both C and Ne begin to burn radiatively 
in the layers above and then transition into convection, expanding the 
material and decreasing the compactness. Once the C and Ne shells 
merge, the resulting convection zone experiences a large dredge-up into 
Ne ash, dropping its convective boundary down to $\approx 2.03\Msun$ 
and creating the large drop in evolutionary compactness.  

For the \Si~simulations, as \fcbm~increases, the evolutionary compactness 
decreases, spanning a range of 0.12 to 0.27. The ONe core mass for these 
simulations is plotted in Figure~\ref{fig:3.3.1_cm} and shows that the depth 
of the dredge-up increases with increasing \fcbm. As the CBM increases, 
the dredge-ups of the C shell overlying the Si burning core deepen in mass, 
mixing in more material and decreasing the size of the ONe core 
(Figure~\ref{fig:3.3.1}, \ref{fig:3.3.1_cm}). All of these simulations have ONe 
core boundaries that correspond to the bottom of the convective shell\footnote{
    Initially these are C burning shells but later in the evolution, they can mix in 
    C and Ne ash and begin to burn C, Ne and O.
    } above this boundary, all of which are below $2.5\Msun$. Similar to the \C\ 
and \ONe\ simulations, the \Si\ simulations with a lower bottom convective 
boundary have smaller compactness.

Although the large changes in the evolutionary compactness for the \C, \ONe\ 
and \Si\ simulations are dominated by the convective C shell growth 
around $2.5\Msun$, decreasing this mass to avoid including these convective
shell interactions in the calculation of $\xi_{2.5}$ would not change the 
non-monotonicity found in the values. The values of $\xi_{2.5}$ change due 
to the expansions and contraction of the core caused by the interaction of 
burning regions within the core. If the mass used in the compactness where 
decreased to avoid the C shell entrainment, the O shells, for example, may 
act in a similar way as the C shells.

Changes to the \fcbm~parameter effect the value of $\xi_{2.5}$, which spans a 
range of $0.12 \leq \xi_{2.5} \leq 0.35$ for the simulations studied here. From 
the ZAMS mass-$\xi_{2.5}$ relation of \citet{Sukhbold2014}, the ZAMS mass 
of $25 \mathrm{M}_{\odot}$ lies at the edge of a relative maximum 
(see \citet{Sukhbold2017} for updated models). These maximums in the mass 
evolution of $\xi_{2.5}$ have been given the name, \textit{islands of 
non-explodability} \citep{Sukhbold2014}. Because the ZAMS mass of the 
stellar models studied here are on the edge of one of these islands of 
non-explodability, and the variation in $\xi_{2.5}$ with respect to \fcbm~is 
similar to the height of the island of non-explodability near this mass, 
some of the variation found in $\xi_{2.5}$ with respect to CBM may be due to 
changes in the ONe core mass mimicking a different ZAMS mass. 

\subsection{Cases for 3D hydrodynamics Simulations}\label{sec:3D hydro}

\begin{figure}
	\includegraphics[width=\linewidth, clip=true, trim=0mm 0mm 0mm 0mm]{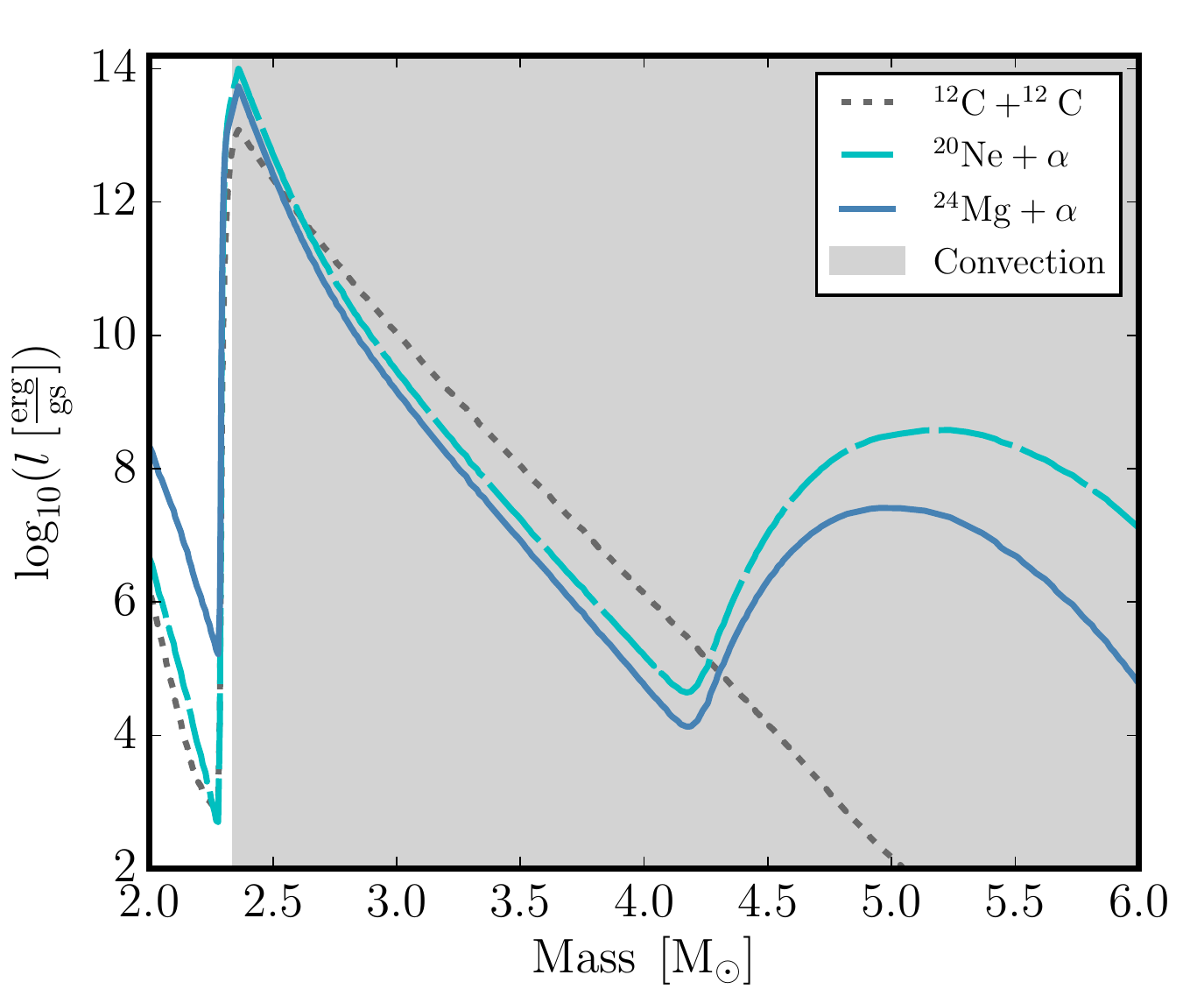}
	\includegraphics[width=\linewidth, clip=true, trim=0mm 0mm 0mm 0mm]{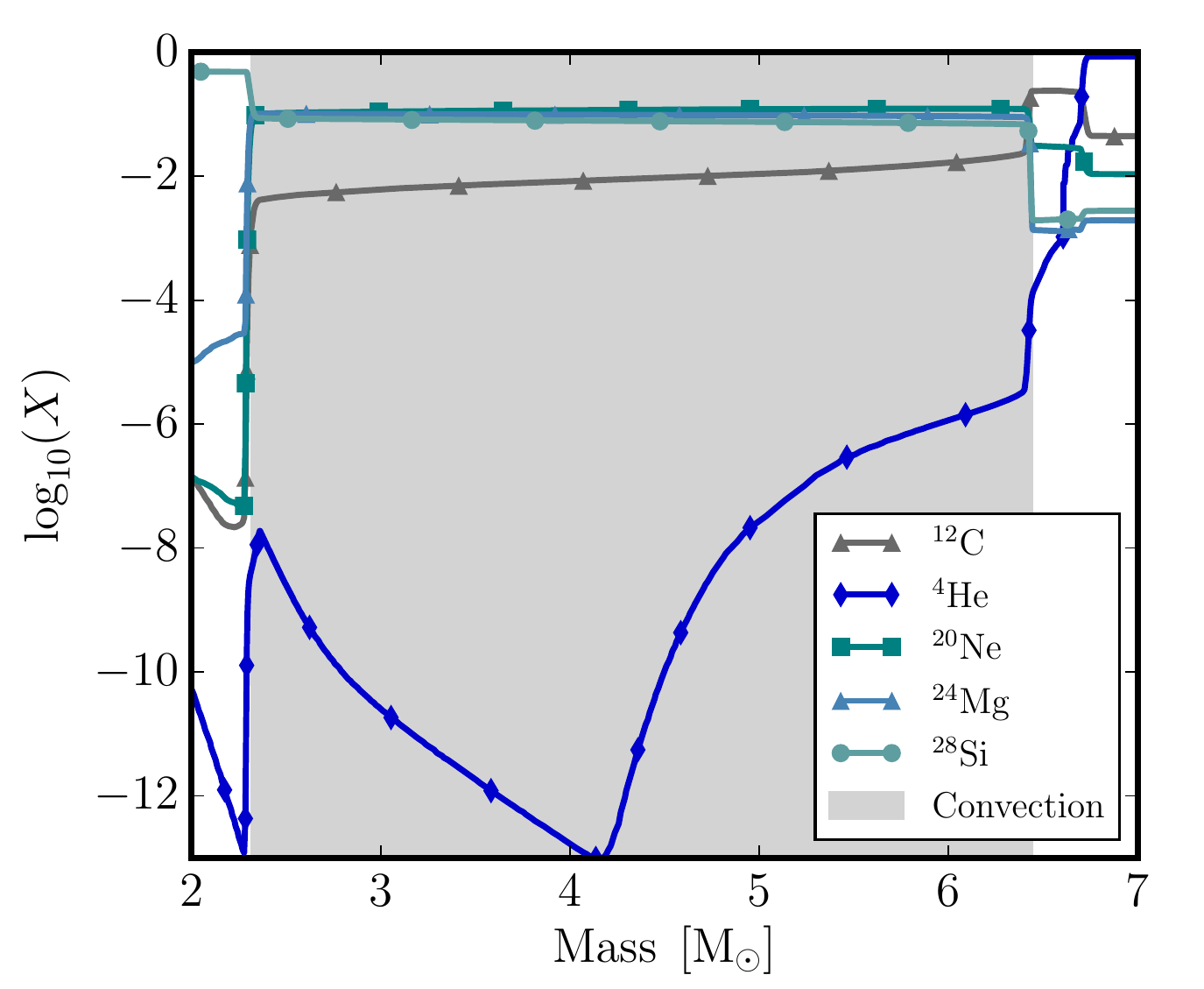}%
	\caption{
		Specific luminosity and abundance profile for the \Cc\ simulation 
		after the merger at $\mathrm{log}_{10}(\tau - t) = -3.47$ (solid 
		black line in Figure~\ref{fig:3.1.2_kip}). The luminosity profile 
		shows the two energy generation regions in the C shell of the 
		\Cc~simulation at this time. The abundance profile shows the 
		distribution of \helium\ in the convective C shell.
	}
	\label{fig:3.1.2_L_abu} 
\end{figure}

\begin{figure}
	\includegraphics[width=\linewidth, clip=true, trim=0mm 0mm 0mm 0mm]{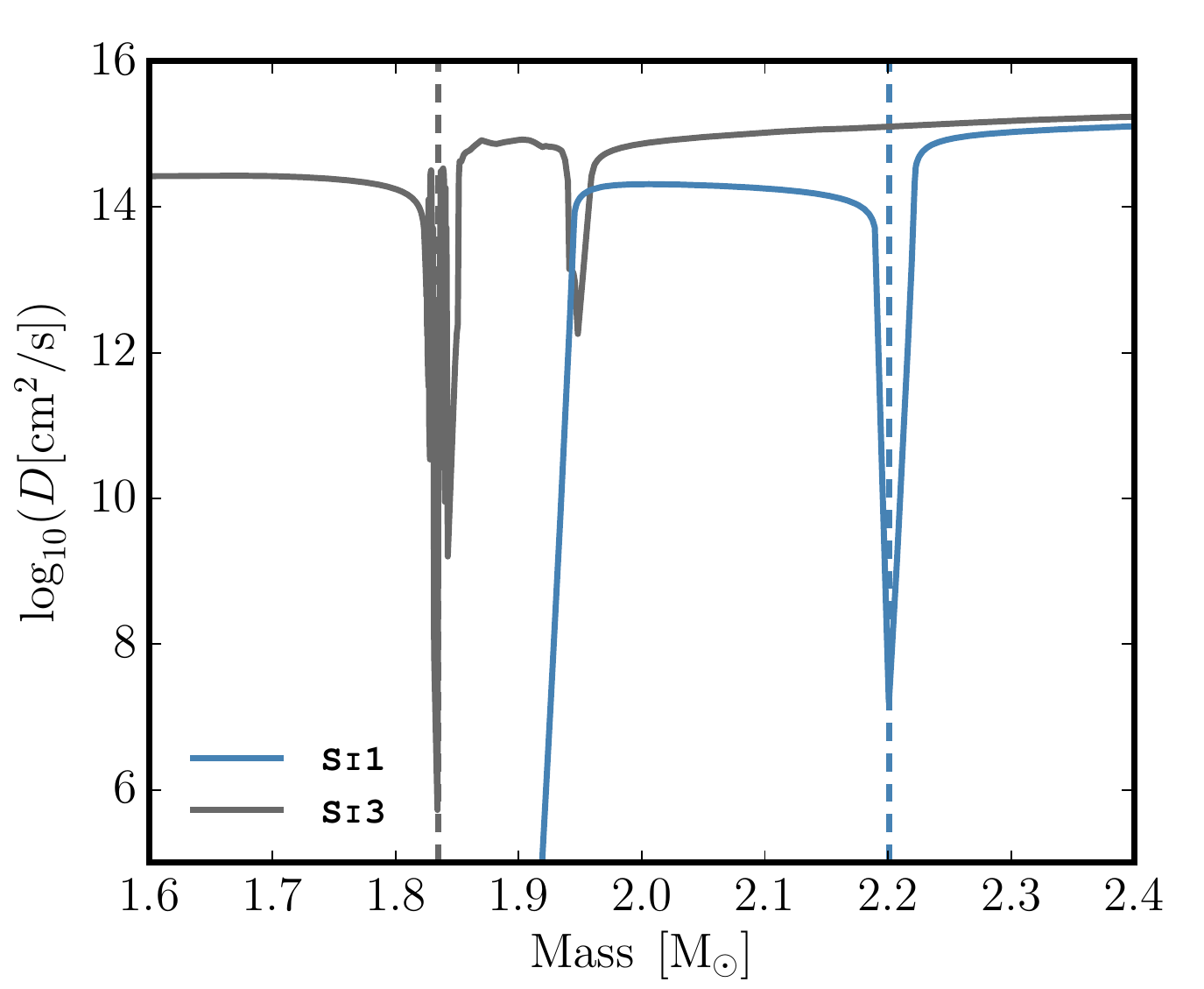}
	\caption{
		The diffusion coefficient used by \mesa~for the \Sia\ and \Sic\ 
		simulations for the bottom of the C shell. The profiles are taken 
		at the black dashed lines in Figure~\ref{fig:3.3.1}. The dotted 
		lines are the convective boundaries between the convection zones. 
		The \Sib\ simulation was excluded for clarity as the diffusion 
		coefficient around this boundary becomes noisy similar to the 
		boundary for the \Sic\ simulation.
	}
	\label{fig:3.3.1_logD} 
\end{figure}

Although the 1D simulations presented here give the cumulative effects of CBM
during the evolution, specific situations occur where 3D hydrodynamic
simulations are necessary in order to understand the mixing. After the shell
mergers and dredge-ups in the \C\ simulations, $\helium$ has a minimum
abundance in the centre of the convection zones. Entropy differences prevent
the merger of shells in the \Si\ simulations although the CBM model allows
some material to mix. Late time shell mergers can occur during the last few
hours of the simulations, these may not affect the elemental distribution
within the simulations as there is not enough time to do so. All of these
events are highly dependent on the fluid dynamics that would be present 
and the 1D simulations may not capture these effects accurately.  

The \C\ simulations all experience two significant burning regions in the
large convective C shells, one at the top and the other at the bottom
(Figure~\ref{fig:3.1.2_kip}). Figure~\ref{fig:3.1.2_L_abu} shows the specific
luminosity from $\carbon+\carbon$, $\neon + \alpha$ and $\magnesium + \alpha$.
Peak energy generation at the top comes from \neon\ and \magnesium\ $\alpha$
capture, as \helium\ can be mixed into the convection zone from above. Deeper
in the convection zone, $\carbon+\carbon$ dominates the energy production from
about 2.5\Msun\ to 4.2\Msun. The C burning produces \helium\ in this
region and this \helium\ is used in further $\alpha$ capture on \neon\ and
\magnesium\ at the bottom of the convection zone. Although the profiles for
\neon\ and \magnesium\ are fairly flat in Figure~\ref{fig:3.1.2_L_abu}, the
profile of \helium\ has a minimum in the convection zone. Because the
\helium\ profile is not flat, the time scale in which \helium\ is being consumed 
and produced is much faster then the convective turn over time scale,
which is about $5 \mathrm{hr}$. This implies that the mixing assumptions 
of MLT might not be valid. In 3D, a convective flow such as this would 
need to be treated as a convective reactive flow, where the fluid dynamics 
are coupled to the reactions. This is necessary because the local turbulent 
mixing in a given region would determine the concentration of each 
species and the energy from the reactions would feed back into the 
fluid dynamics.

During Si burning the Ne, O and Si convective shells form relatively close to
each other in mass. Because the entropy gradients are smaller in this region
than further out, there is a potential for nearby convective shells to merge.
In the \Sia, \Sib\ and \Sic\ simulations, after convective core Si burning,
an O shell forms (dashed line in Figure~\ref{fig:3.3.1}) at 
$\mathrm{log}_{10}(\tau - t) \approx -3.5$. The \template\ simulation forms 
an O shell later on in its evolution. In the \Sia\ and \Sib\ simulations, a 
convective Ne shell exists above the O shell during its formation. The three 
convective shells are only separated from the convective C shell by a very 
small mass, in some cases, $< 0.05\mathrm{M}_{\odot}$. Despite the 
proximity and enhanced CBM of these shells, they do not merge 
(Figure~\ref{fig:3.3.1_logD}). Strong entropy gradients between these 
shells inhibit the mixing across the shell boundaries and prevent
the shells from merging (Figure~\ref{fig:3.3.1_s}). Although these shells don't
merge, because of the enhanced CBM of the \Si\ simulations and the small
separation of the convective shells, some mixing still occurs between them. The
exponential decay of the diffusion coefficient across the boundary allows for a
region outside of the convection zone to mix with material from the convection
zone. In this case, both convection zones are mixing into this small radiative
region separating them (Figure~\ref{fig:3.3.1_logD}). This configuration may 
be analogous to a double convective boundary, where the boundary between 
two convection zone is stable.

During Si and Fe core burning, C, Ne and O shells can exist separated by 
relatively small masses (Figure~\ref{fig:3.3.1}). Specifically, in the \Sib\ 
simulation, a convective Si shell approaches an O shell underlying the 
convective C shell (black solid line in Figure~\ref{fig:3.3.1}). These shells 
are in the process of merging, although a relatively slow mixing event such 
as this (compared to the nuclear time scale) would have little effect on the 
abundances in the C shell. This is because the convective turn over time 
scale of the C shell at this time is $\approx 5 \mathrm{hr}$, and the star 
has roughly the same amount of time left before collapse. Therefore, any 
mixing event that didn't significantly decrease the convective turn over 
time scale of the C shell may not be able to mix material high enough
in time to surpass the fallback mass. The results given by the 1D 
\mesa~simulations do not model this event appropriately, as the mixing 
between the shells is most likely dependent on the fluid dynamics of 
convective mixing.

All of these situations are dependent on the dynamics of turbulent convective mixing between fluids over relatively short time scales, which can't be captured by the diffusive approximation. The production and consumption of $\helium$ in the in the C shells of the \C\ simulations is much faster then the convective flows within the shell. The CBM model allows for some mixing between convective shells that have not merged, which is similar to two convective shells separated by a stable boundary in 3D. The late time mixing between the shells which have convective turnover time scales on the order of the evolutionary life time may effect the final abundances. In order to understand how the mixing is happening in these situations, 3D hydrodynamic simulations are necessary. 

\section{Summary and Discussion} \label{sec: summary}


In order to test the effects of CBM on the post He core evolution of a 25\Msun,
solar metallicity stellar model, the CBM parameter, \fcbm, was varied at
different evolutionary times (see Section~\ref{sec: methods}, 
Figure~\ref{fig:2_kip_runs}). The simulations implemented values of \fcbm~of
0.002, 0.012, 0.022 and 0.032 with the \template\ simulation having the lowest
value, acting as a control for the other simulations (Figure~\ref{fig:2_runs}).
The simulations were run with $\approx 3000$ spatial zones and $\approx
300,000$ time steps.

Enhanced CBM promotes dredge-ups and shell mergers which restructure 
the core significantly leading up to the end of the star's life. The effects of 
enhanced CBM on a single convection zone can be seen in the first C shell 
of the \C\ simulations (Section~\ref{sec: Convective structure}). CBM 
decreases the lifetime of the first C shell and mixes ash into the convection 
zone from below. The ash distribution left behind by the first C shell defines 
the starting point for the second C shell. Enhanced CBM also pushes the 
bottom convective boundary deeper in mass into the ONe core, burning C 
deeper in the star and mixing up C ash. The net effect is similar compared to the He-shell flash convection in AGB stars, in which CBM at the bottom of the convection zone drives the Lagrangian coordinate of the bottom boundary deeper into the underlying core and adding material from below to the He-shell flash convection \citep{Herwig2000}. In that situation CBM can accommodate several observational properties of AGB \citep{Herwig2005} and post-AGB \citep{Werner06} stars.  The situations is also reminiscent of the effect of CBM in simulations of nova \citep{Denissenkov2012} where it causes models to have a fast rise time and enhancements of C and O in the ejecta, as observed.

The effect of CBM on the convection 
zones compound during Ne, O and Si core burning for both the \C\ and 
\ONe\ simulations. In these simulations enhanced CBM interacts with the 
C, Ne and O shells by promoting dredge-ups and shell mergers, acting to 
restructure them. The dredge-ups of the C shells are of particular significance 
(Section~\ref{sec:Dups}).  During the growth of a new convective C shell, 
dredge-ups can mix Ne ash from radiative Ne burning into the C shell, 
decreasing the location in mass of the bottom boundary. These mixing 
events can transport Ne ash found under the C shells to the tops of the 
convection zones, provided there is sufficient time to do so before the end 
of the stars life. Similarly, in the \Cc\ simulation, a convective Ne shell 
merges with the newly formed C shell having a similar effect. Because 
these mixing events happen around $3.5 \mathrm{d}$ before collapse, 
the material can be mixed throughout the C shell, as the convective turn over 
time scale of the C shell is $\approx 5 \mathrm{hr}$. In contrast, the C shells 
of the \template\ simulation do not experience any significant dredge-up or 
shell merger with the underlying material. 


Using the fallback prescription given by \citet{Fryer2012} with a mass cut of
$M_{\mathrm{delay}}=5.7\Msun$, the presupernova overproduction factors, $\Theta_{i}$, 
from these simulations show large variations in C, Ne, Mg and Si 
(Section~\ref{sec: profac}). The largest deviations from the \template\ occur 
in simulations that experience enhanced CBM during C, Ne and O core 
burning (\C\ and \ONe) due to the dredge-ups and shell mergers that occur 
in those simulations. These mixing events happen early enough in the 
evolution of the core that the Ne and O ash can be mixed to the top of the C 
shell, above the mass cut. Although the \Si\ simulations have tightly packed 
convective shells near the end of their evolution (Section~\ref{sec:3D hydro}), 
they do not merge due to the large entropy gradients between them. Changes 
in $\Theta_{i}$ for these simulations are dominated by the C shell dredge-ups 
found earlier in the evolution. With a convective turnover time scale of 
$\approx 5 \mathrm{hr}$ during Si shell burning, the C shell needs a 
large increase in luminosity to decrease the convective turnover time scale 
to a value less then the time left till collapse ($\approx 30 \mathrm{min}$). 
Dynamic events such as dredge-ups and shell mergers in the late time of 
evolution could potentially provide the luminosity, but events energetic 
enough to do this are not seen in the simulations at this time. 

The mass cut, $M_{\mathrm{delay}}=5.7\Msun$, lies under the upper 
convective boundary of the C shell for all simulations. It is 
$\approx 1.5 \mathrm{M}_{\odot}$ 
from the top of the C shell and $\approx 3.5 \mathrm{M}_{\odot}$ from the 
bottom for the \template. The C shell is convective near collapse for these 
simulations and 
the abundances of the elements plotted Figure~\ref{fig:profac} are fairly 
mixed when the shell is not experiencing a dredge-up or shell merger. 
Therefor variations to the mass cut on the order of a solar mass or more do 
not have a large effect on the $\Theta_{i}$ distribution. The amount of ejected 
material changes but the relative quantities are insensitive to variations of 
the mass cut within the C shell. If the mass cut were to vary within this mass 
range, in order for $\Theta_{i}$ to change, material from below will still need to 
be mixed into the C shell. The mass cut of $M_{\mathrm{delay}}=5.7\Msun$ 
was taken as an approximation in order to determine the $\Theta_{i}$ values for 
the simulations and illustrate the nucleosynthetic effects of CBM mechanisms 
on the late stage structure. The overproduction factors assume 
different evolutionary outcomes than that given by the compactness. For 
example, a simulation that collapses to a black hole as determined by the 
compactness, should, but will not have an overproduction 
factor showing zero mass ejected. These two diagnostics are meant to be 
used to investigate the CBM in the stellar cores rather than provide 
nucleosynthetic yields or determine the explodability of a particular model.  


The compactness of each simulation is dependent upon the strength of the CBM
(Section~\ref{sec: compactness}). The values of $\xi_{2.5}$ range from 0.12 to
0.35 when taken 30 seconds before collapse. Simulations which experience
enhanced CBM during the Ne and O core burning stages (\C~and \ONe) show 
non-monotonicity of $\xi_{2.5}$ with respect to \fcbm, whereas
$\xi_{2.5}$ is a monotonically decreasing function of \fcbm~in the
\Si~simulations.  Both the \C~and \ONe~sets diverge from the
\template~significantly during C and O burning as dredge-ups and shell mergers 
change the CO core structure, creating different numbers of convective shells. 
This means that in these simulations, the non-monotonicity in
$\xi_{2.5}$ primarily comes from the C, Ne and O core and shell interactions in
the form of dredge-ups and shell mergers during those core burning stages, but
not during the end of Si core and shell burning. The ONe core mass of the
\Si~simulations decrease with increasing CBM and do not show the non-monotonic
deviation for the \template. During the late stages of evolution in the
\Si~simulations, the main effect of the CBM is to decrease the bottom C shell
boundary. Because this boundary is below $2.5\Msun$ the values
of $\xi_{2.5}$ represent the amount of material that is in the C shell rather
than any intricate shell interactions. This means that the variation found in $\xi_{2.5}$ 
is due to the interaction of the C, Ne and O convective regions during those core 
burning stages where as during Si and Fe core, the mixing events have little effect. 
This is mainly because the time scales over which
these mixing mechanisms can change the structure are longer than the time
remaining until collapse. This result is somewhat consistent with \citet{Sukhbold2017}
who found that the non-monotonic behaviour of $\xi_{2.5}$ with respect to ZAMS
mass around the islands of non-explodability is mainly due to the formation of
the C and O burning shells. 
 
The variation in $\xi_{2.5}$ of $\Delta \xi_{2.5} = 0.23$ is due to two
simulations with the highest values of \fcbm~(\Sic~and \Cc). If only the
simulations implementing \fcbm~values of 0.012 and 0.022 are considered, the
variation in $\xi_{2.5}$ decreases to $0.15$, but the non-monotonicity
found in the evolutionary compactness is still present. This means that the
impact of CBM on the compactness is a cumulative effect of convective shell
interactions and is not just limited to the cases where large amounts of mixing
change the structure.

The progenitor structures for all models presented in this work are available
for download \red{HERE (link and DOI to appear)}.

\section*{Acknowledgements}
SJ is a Director's Fellow at Los Alamos National Laboratory and acknowledges
prior support from the Alexander von Humboldt Foundation and Klaus Tschira
Foundation (KTS). FH acknowledges funding from an NSERC Discovery grant. The simulations were performed on the Compute Canada/WestGrid cloud systems Arbutus at the University of Victoria. 
 
\bibliography{CBM_BIB}  

\label{lastpage}
\end{document}